\documentclass[12pt,preprint]{aastex}
\usepackage{epsfig}
\usepackage{rotating}
\usepackage{natbib}
\def\etal{{et al.\thinspace}}

\def\spose#1{\hbox to 0pt{#1\hss}}

\def\multleft#1{\hbox to size{\vbox {\halign {\lft{##}\cr #1}}\hfill}\par}
\def\multright#1{\hbox to size{\vbox {\halign {\rt{##}\cr #1}}\hfill}\par}

\def\degmark{^\circ}
\def\boxit#1{\vbox{\hrule\hbox{\vrule\kern3pt\vbox{\kern3pt
          #1 \kern3pt}\kern3pt\vrule}\hrule}}

\def\cm{{\rm\thinspace cm}}

\def\erg{{\rm\thinspace erg}}
\def\eV{{\rm\thinspace eV}}

\def\K{{\rm\thinspace K}}
\def\keV{{\rm\thinspace keV}}
\def\km{{\rm\thinspace km}}

\def\Mpc{{\rm\thinspace Mpc}}
\def\Msun{\hbox{$\rm\thinspace M_{\odot}$}}

\def\ph{{\rm\thinspace ph}}
\def\s{{\rm\thinspace s}}
\def\ks{{\rm\thinspace ks}}

\def\cts{{\rm\thinspace cts}}

\def\ergcmps{\hbox{$\erg\cm\s^{-1}\,$}}
\def\ergpcmps{\hbox{$\erg\cm^{-1}\s^{-1}\,$}}
\def\ergpcmsqps{\hbox{$\erg\cm^{-2}\s^{-1}\,$}}

\def\ergps{\hbox{$\erg\s^{-1}\,$}}

\def\kmps{\hbox{$\km\s^{-1}\,$}}
\def\kmpspmpc{\hbox{$\km\s^{-1}\Mpc^{-1}\,$}}

\def\pcmsq{\hbox{$\cm^{-2}\,$}}

\def\phpcmsqps{\hbox{$\ph\cm^{-2}\s^{-1}\,$}}

\def\ctsps{\hbox{$\cts\s^{-1}$}}

\makeatletter

\let\@internalcite\cite
\def\cite{\@ifstar{\citey}{\citefull}}
\def\citefull{\def\astroncite##1##2{##1\ ##2}\@internalcite}
\def\citey{\def\astroncite##1##2{##1\ (##2)}\@internalcite}
\def\citeyear{\def\astroncite##1##2{##2}\@internalcite}
\def\citename{\def\astroncite##1##2{##1}\@internalcite}
\def\@citex[#1]#2{\if@filesw\immediate\write\@auxout{\string\citation{#2}}\fi
  \def\@citea{}\@cite{\@for\@citeb:=#2\do
    {\@citea\def\@citea{; }\@ifundefined
       {b@\@citeb}{{\bf ??}\@warning
       {Citation `\@citeb' on page \thepage \space undefined}}%
{\csname b@\@citeb\endcsname}}}{#1}}
 
\def\@cite#1#2{#1\if@tempswa #2\fi} 
\def\@biblabel#1{}
 
\def\astroncite#1#2{#1\ #2}
\makeatother

\begin{document}

\title{Spectral Analysis of the Accretion Flow in NGC~1052 with Suzaku}

\author{L.~W.~Brenneman\altaffilmark{1}, 
K.~A.~Weaver\altaffilmark{2},
M.~Kadler\altaffilmark{3,4,5},
J.~Tueller\altaffilmark{2},
A.~Marscher\altaffilmark{6},
E.~Ros\altaffilmark{7},
A.~Zensus\altaffilmark{7},
Y.Y.~Kovalev\altaffilmark{7,8},
M.~Aller\altaffilmark{9},
H.~Aller\altaffilmark{9},
J.~Irwin\altaffilmark{9},
J.~Kerp\altaffilmark{10},
S.~Kaufmann \altaffilmark{11}}

\altaffiltext{1}{NPP Postdoctoral Fellow (ORAU); NASA's GSFC, mail code 662, Greenbelt MD~20771~USA}
\altaffiltext{2}{NASA's GSFC, mail code 660, Greenbelt MD~20771~USA}
\altaffiltext{3}{Dr. Karl-Remeis-Sternwarte, Astronomisches Institut
  der Universit\"{a}t Erlangen-N\"{u}rnberg, Sternwartstr. 7, 96049
  Bamberg, Germany}
\altaffiltext{4}{CRESST/NASA Goddard Space Flight Center, mail code 662
  Greenbelt, MD~20771~USA}
\altaffiltext{5}{Universities Space Research Association, 10211
  Wincopin Circle, Suite 500 Columbia, MD~21044~USA}
\altaffiltext{6}{Institute for Astrophysical Research, Boston University, 725 Commonwealth Ave., Boston, MA~02215~USA}
\altaffiltext{7}{MPIfR, Postfach 2024, D-53010 Bonn, Germany}
\altaffiltext{8}{Astro Space Center of Lebedev Physical Institute,
  Profsoyuznaya 84/32, 117997 Moscow, Russia}
\altaffiltext{9}{Department of Astronomy, University of Michigan, 500
  Church St., Ann Arbor, MI~48109~USA}
\altaffiltext{10}{Argelander-Institut f\"{u}r Astronomie, Universit\"{a}t Bonn,
  Auf dem H\"{u}gel 71, D-53121 Bonn, Germany}
\altaffiltext{11}{Landessternwarte, Universit\"{a}t Heidelberg,
  K\"{o}nigstuhl 12, D-69117 Heidelberg, Germany}

\begin{abstract}
We present an analysis of the $101 \ks$, 2007 {\it Suzaku} spectrum of
the LINER galaxy NGC~1052.  The $0.5-10 \keV$ continuum is
well-modeled by a power-law modified by Galactic and
intrinsic absorption, and it exhibits a soft, thermal emission component
below $1 \keV$.  Both a narrow core and a broader component of
Fe K$\alpha$ emission centered at $6.4 \keV$ are robustly detected.  While the
narrow line is consistent with an origin in material distant from the
black hole, the
broad line is best fit empirically by a model that describes
fluorescent emission from the inner accretion disk
around a rapidly rotating black hole.  We find no evidence in this
observation for
Comptonized reflection of the hard X-ray source by the disk above $10
\keV$, however, which casts doubt on the hypothesis that the broad
iron line originates in the inner regions of a standard accretion disk.  We explore other possible
scenarios for producing
this spectral feature and conclude that the high equivalent width ($EW
\sim 185 \keV$) and
full-width-half-maximum velocity of the broad iron line ($v \geq 0.37c$) necessitate an
origin within $d \sim 8\,r_{\rm g}$ of the hard X-ray source.  Based
on the confirmed presence of a strong radio jet in this galaxy nucleus, the broad
iron line may be produced in dense plasma near the base of the jet,
implying that emission mechanisms in the centralmost portions of
active galactic nuclei are more complex than previously thought.
\end{abstract}


\section{Introduction}
\label{sec:intro}

Since their first identification as an interesting class at optical wavelengths
\citep{Heckman1980}, the nature of Low Ionization Nuclear Emission
Regions (LINERs) has remained elusive.  Many such regions are
associated with active galactic nuclei (AGN), though their X-ray luminosities are highly
sub-Eddington and significantly below those of their Seyfert galaxy counterparts.  The
reason for this modest nuclear energy output remains unknown: possibilities
include heavy absorption of the nuclear emission, comparably inefficient radiation of the
accretion flow, or some combination of both.  

The notion of an
advection-dominated accretion flow (ADAF) may apply in LINERs, wherein the ions
lose thermal contact with the electrons and only a small fraction of
the dissipated energy is radiated \citep{Rees1982,Narayan1994}.  Much
controversy remains as to whether this picture is correct, however, and the
debate has become more pointed since the report by \citet{Ho1997d}
that some $20-30\%$ of all galaxies are members of the LINER class.

X-ray spectroscopy can help in determining whether
the accretion disk is in an ADAF state or not.  A LINER galaxy harboring an AGN
produces a characteristic hard X-ray continuum.  If the accretion flow
close to the black hole conforms to the conventional 
thin disk archetype \citep{Shakura1973}, the cool gas orbiting in the
inner disk should
intercept and reprocess a significant fraction of the continuum, also
producing the Fe K$\alpha$ line (among many fluorescent emission lines from other
species as well).  This spectral feature should have a broadened,
skewed profile created by the relativistic properties of the
spacetime close to the black hole, and a significant Comptonized reflection
component should be seen in the continuum above $10 \keV$ as well.
If, however, the accretion flow
in this region takes the form of an ADAF, the ion temperature is so
high that the iron ions should be totally stripped of electrons.
Furthermore, the trapped energy will puff up the disk, reducing its
density and optical depth.  Both of these conditions render the
reflected emission, including the
emission of Fe K$\alpha$, virtually non-existent in this region.  In
this case, only contributions to the Fe K$\alpha$ line from
other locations (e.g., from a disk wind, the outer disk, broad line region or putative
molecular torus) will
manifest, and the line may likely have a much narrower profile as a
result.

The LINER galaxy NGC~1052 is an interesting case in which to study the
dynamics of the accretion flow.  
The AGN is housed by a nearby elliptical host galaxy with a redshift of
$z=0.0049$ \citep{Knapp1978},
implying a distance of $20.7 \Mpc$ using WMAP cosmology.
The source has a well-studied LINER optical
spectrum (e.g., \citet{Gabel2000}), and is also reasonably bright in the X-ray band
for a LINER, with an average $2-10 \keV$ flux of $F_{2-10} \sim 5 \times 10^{-12}
\ergpcmsqps$.  
Using the $M-\sigma$ relation of \citet{Tremaine2002} and the stellar
velocity dispersion of \citet{NW1995}, \citet{Woo2002} derived a black
hole mass of $M_{\rm BH}=1.54 \times 10^8 \Msun$ for NGC~1052, and a
bolometric luminosity of $L_{\rm bol}=6.92 \times 10^{43} \ergps$
using flux integration over its spectral energy distribution.
Combining these results with the derived Eddington luminosity yields
an estimated accretion efficiency for the black hole in NGC~1052 of $\eta
\leq 0.004$, in Eddington units.  This is consistent with
general results from LINERs.
The source is known to be heavily absorbed based on the
presence of ${\rm H}_2{\rm O}$ megamasers in its core
\citep{Braatz1994}, which lie along the direction of a radio jet also
present in the galaxy \citep{Claussen1998}.  Recent VLBI studies have
constrained the jet inclination angle to lie between $57-72\degmark$ to
our line of sight \citep{Kadler2004a}, implying that the jet emission
is de-beamed, and suggesting that the bulk of the
X-ray emission is not produced by the jet.   

The observed X-ray spectrum of NGC~1052 is quite flat (for the power-law
component of the emission $\Gamma \sim 0.2-1.7$,
depending on the data and models used) and appears to be dominated at
lower energies by thermal and/or photoionized emission
\citep{Weaver1999,Guainazzi2000,Kadler2004b}.  The
source of this soft emission is likely the interaction of the jet with the
surrounding ISM \citep{Kadler2004b}, while the power-law component at
harder energies is thought to originate from nuclear emission.  Radio
continuum and {\it RXTE} monitoring both show that the
source varies strongly on time scales of weeks to months.  The $2-10
\keV$ flux, in particular, varies between $\sim 4-9 \times 10^{-12}
\ergpcmsqps$, and though there are large
uncertainties in the {\it RXTE} data, the spectrum appears to switch
from softer to harder states,
corresponding to changes in overall source flux (Kadler \etal, in prep.).

A prominent Fe K$\alpha$ line has been previously observed in the source by {\it ASCA}
\citep{Weaver1999}, {\it BeppoSAX} \citep{Guainazzi2000}, {\it
  XMM-Newton} (Kadler \etal, in prep.), and {\it Chandra}
\citep{Kadler2004a,Ros2007}.  Of these observations, a broadened
component to the Fe K$\alpha$ line was seen with {\it BeppoSAX} and {\it
  XMM-Newton}.  Given the low accretion rate of
NGC~1052 --- a scenario which has long been thought to produce ADAFs,
e.g., \citet{Narayan1994} --- it is natural to speculate on the origin
of the broad iron line.  The spectral state of NGC~1052 is in many
ways analogous to the low/hard state observed in X-ray binaries
(XRBs).  Contrary to the ADAF paradigm, \citet{Miller2006} reported
that {\it XMM-Newton} observations of the relativistically broad Fe
K$\alpha$ line in the low/hard state of XRB GX 339--4 showed a
standard thin disk remaining at or near to the innermost stable
circular orbit (ISCO), at least in bright phases of the low/hard
state.  The authors therefore argued that potential
links between the inner disk radius and the onset of a steady compact
jet, and the paradigm of a radially recessed disk in the low/hard
state, do not hold universally.  Note, however, that
\citet{Gierlinski2008} cautioned against using only the shape of the Fe K$\alpha$
line to determine the radius of the ISCO, suggesting that these results can be
significantly underestimated unless 
irradiation from the energetically dominant hot plasma (along with
other spectral effects) is taken into account. 
If the broad iron line in NGC~1052 does originate from the
accretion disk, however, then this
spectral feature, coupled with detailed
VLBI observations of the inner jet
\citep{Kadler2004b,Kadler2004a}, may offer
a unique opportunity to study the connection between the
accretion flow and jet production in AGN.   

A link between accretion and jet activity in AGN has been
noted in the radio galaxies 3C~120
\citep{Kataoka2007,Marscher2006,Marscher2002} and 3C~111
\citep{Marscher2006}.  While Marscher \etal have noted dips in the
X-ray light curves of these sources correlated with ejections of
superluminal knots in the radio jet, suggesting that a portion of the
inner disk is disturbed during an ejection event, these results fall
short of mapping out the physical structure of the disk at the time.
However, Kataoka \etal
observed a broad Fe K$\alpha$ line in 3C~120 with {\it Suzaku} in both
the high and low flux states of the source.  This discovery
demonstrates the viability of studying the accretion flow and the jet
activity of radio-loud AGN simultaneously, and will hopefully pave the
way for new insights into the fundamental link between these two
processes.  

Here we discuss the analysis of our 2007 {\it Suzaku} observation of
NGC~1052, which offers the highest quality X-ray spectrum of this galaxy
to date and gives us an unprecedented, simultaneous view of the $0.5-35 \keV$ energy
range.  NGC~1052 is one of a small handful of AGN
sufficiently bright in both the X-ray and radio bands for tests of a
jet/disk connection.  Additionally, it is the only source for which
we will have a full package of multi-epoch X-ray spectral data, 
X-ray and radio light curves, and VLBA imaging within a common time
window and with an analysis by a single collaborative group.  
A forthcoming paper will provide full details of this ongoing
campaign, placing our {\it Suzaku} 
results in the context of prior X-ray and ongoing radio analysis of
this source (Kadler \etal, in prep.). 

Our {\it Suzaku} observation of NGC~1052 and the data reduction will be
detailed in Section 2.  We discuss variability over the course of the
observation in \S3.  Our approach to the spectral fitting of the
time-averaged data and the results of this analysis for both the XIS and the HXD/PIN
instruments are presented in \S4.  We examine the Fe K$\alpha$ region
at length in \S4 as well.  A comparison with results from previous
epochs appear in \S5, and we discuss our results in context in \S6.
Our conclusions comprise \S7.

\section{Observations and Data Reduction}
\label{sec:reduction}

We observed NGC~1052 for a total of $\sim 101 \ks$ from
16-18 July 2007 with {\it Suzaku} in the HXD nominal pointing position. 
Reduction of the {\it Suzaku} data followed the procedures outlined in
\S4.7  and \S4.8 of the {\it Suzaku} ABC Guide, available
online\footnote{http://heasarc/docs/suzaku/analysis/abc/node1.html}.
Calibration files used were the latest version as of the time of this
writing (January 2009).  This calibration was incorporated into the
X-ray Imaging Spectrometer (XIS)
data reduction using the {\tt xispi} task on the unfiltered data.
Once cleaned, the event files for the XIS~0, XIS~1 (XIS~2 became defunct
as of Nov. 2006) and XIS~3 detectors were loaded into
{\sc xselect} for reprocessing.  Images were extracted that enabled identification
of source and background regions.  As per the recommendations in the
ABC Guide, we employed a circular source region $\sim 250''$ in
radius, and maximized our background region area while avoiding the
calibration sources at the corners of each detector chip. 
After spatial filtering was
applied, spectra and light curves were extracted for each of the
operational three detectors.  Response matrices and ancillary response
files were generated using the {\tt xisresp} task, and the spectra,
backgrounds and responses for
the front-illuminated chips (XIS~0 and XIS~3) were co-added to increase
the signal-to-noise using {\tt addascaspec}.  The spectra, backgrounds
and responses were then rebinned by a factor of eight (to $512$ channels)
using the {\tt rbnpha} and {\tt rbnrmf} tasks to speed spectral
fitting.  Finally, the spectra, backgrounds and responses were 
grouped together
using the {\tt grppha} task.  Only spectral channels with
with a minimum of $20$ cts/bin were used for fitting in order to
ensure the validity of the chi-squared statistic.  This corresponded
to an energy range of $0.7-9 \keV$ for the XIS~0+3 data ($32113$ total
counts) and $0.5-7 \keV$ for the XIS~1 data ($16146$ total counts).

{\it Suzaku} did observe NGC~1052 with the silicon diode PIN instrument of the
Hard X-ray Detector (HXD), but the GSO crystal scintillator instrument did not have enough counts
to provide useful data.  The PIN data were reduced using the
``tuned'' calibration files for the non-X-ray background (NXB) and the
response files that had been generated for the period surrounding the observation.
These files were downloaded from the {\it Suzaku} Guest Observer
Facility (GOF) for use.  We
first needed to ensure that only the portion of the NXB coincident
with our observation was used in processing, so we merged the
good-time interval (GTI) of the NXB with that of our screened event
file to yield a common GTI using {\tt mgtime}.  With {\sc xselect}, we then filtered
the data and background using this common GTI and extracted our
spectra and light curves for the source and NXB.  We corrected for dead time in
the observation using the {\tt hxddtcor} task, and increased the
exposure time of the NXB data by a factor of ten to compensate for the
inflated event rate, which had been similarly multiplied in an effort
to suppress Poisson noise.  As for the XIS data, only spectral channels with
with a minimum of $20$ cts/bin were used for fitting, which limited
our energy range to $12-35 \keV$.  After reduction and filtering, the
PIN spectrum had $35705$ total counts over this energy range.  

The cosmic X-ray background (CXB) has not
been taken into account in PIN observations, so to model it
appropriately, we simulated the CXB spectrum in {\sc xspec} \citep{Arnaud1996}
using a power-law model
of $\Gamma=1.29$ with a high energy cutoff and foldE$=40 \keV$ (initial
parameters are given in the ABC Guide).  We then folded this model
through the response file for the PIN flat field for an exposure time
of $10^6 \s$ to generate the simulated CXB data.  This spectrum was
then read back into {\sc xspec} with the PIN response for our
observation and the original model was refitted over $12-35 \keV$ to yield the actual
parameters of the CXB during this time.  In our case, the simulated CXB spectrum contributed a
count rate of $\sim 2.28 \pm 0.02 \times 10^{-2} \ctsps$ ($\sim 5\%$)
to the total X-ray
background from $12-35 \keV$.  In modeling the PIN data with {\sc xspec}, we
applied the best-fit CXB parameters as a constant with each fit.  The
CXB followed a power-law form with a high-energy cutoff:
$\Gamma=1.30 \pm 0.19$, $K_{\rm po}=8.13 \pm 3.07 \times 10^{-4} \phpcmsqps$, ${\rm
  foldE}=40.53 \pm 15.93 \keV$.  The PIN instrument detected
the source above the (NXB+CXB) background at a $> 10\sigma$
level in our observation.

\section{Timing Analysis and Variability}
\label{sec:timing}

\begin{figure}
\begin{center}
\includegraphics[width=0.3\textwidth,angle=270]{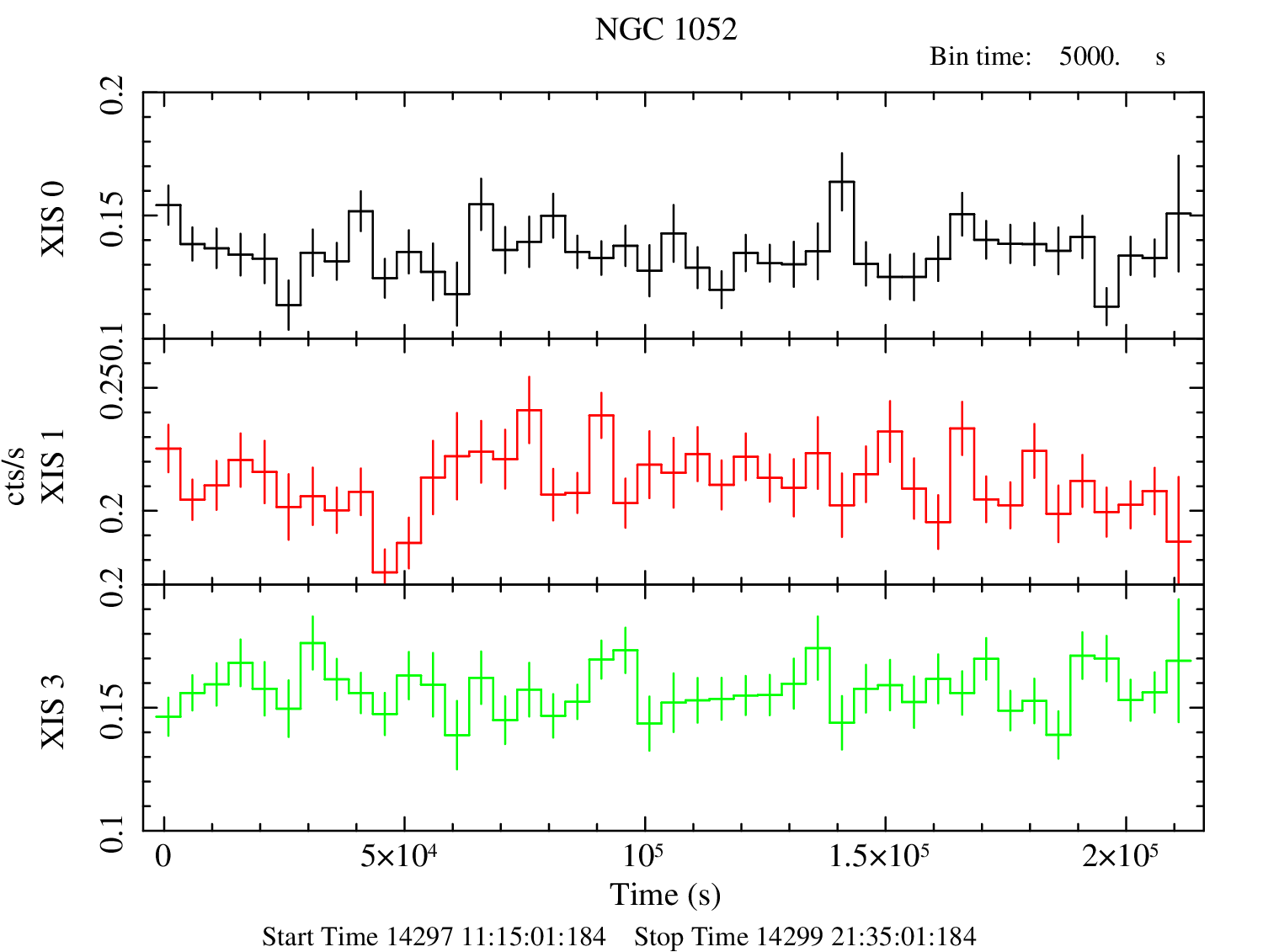}
\includegraphics[width=0.3\textwidth,angle=270]{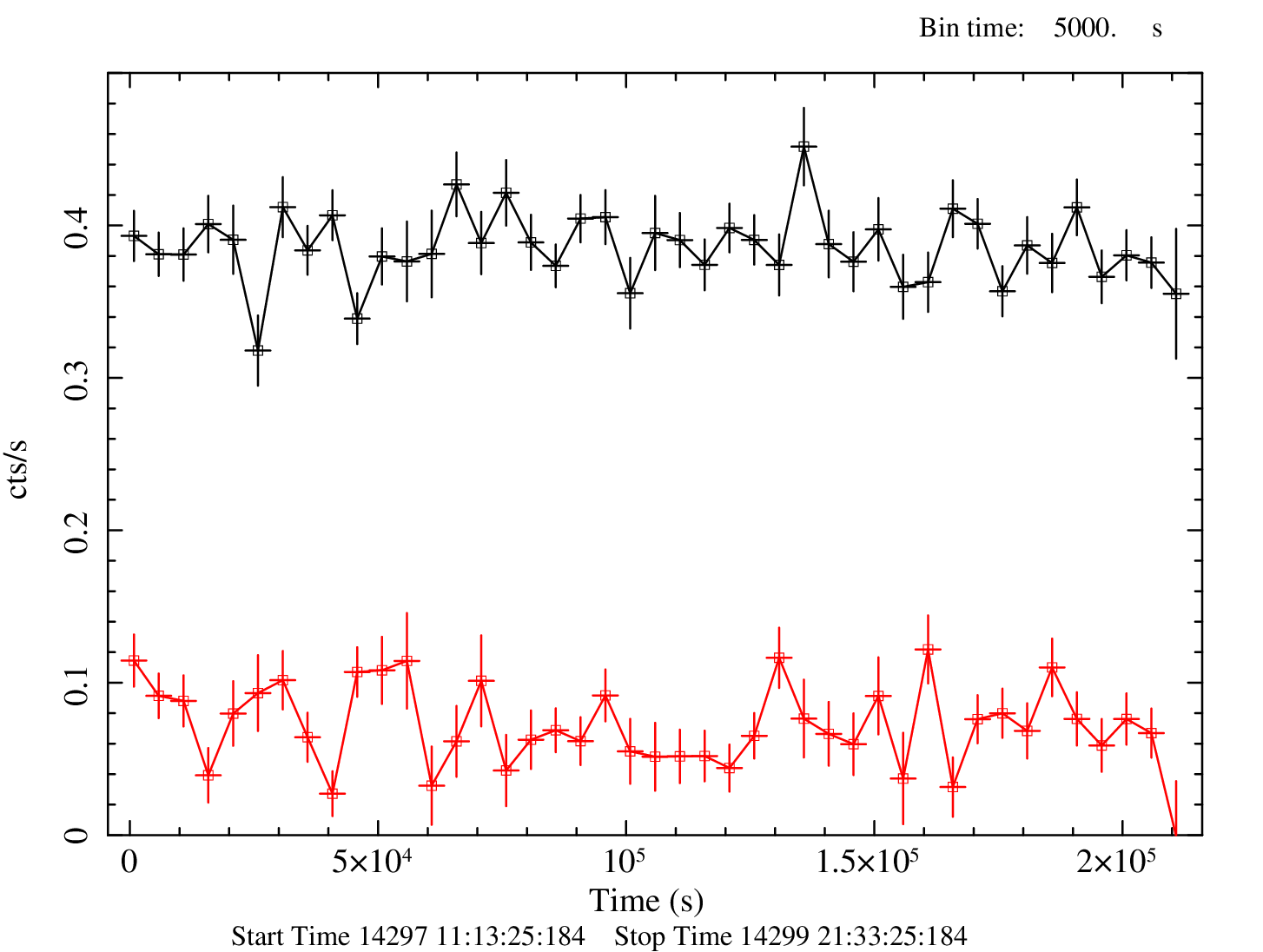}
\end{center}
\caption{\small {\it Left:} Light curves from the three operational XIS
  instruments for the 2007 {\it Suzaku} observation of NGC~1052.  {\it
  Right:} Light curve of the combined XIS data (black) as well as that of the
  PIN (red) over the course of the observation.  All light curves
  shown are background-subtracted, including CXB subtraction for the PIN
  data.  Note the overall lack of significant variability of the source.}
\label{fig:lcurves}
\end{figure}

\begin{figure}
\begin{center}
\includegraphics[width=0.3\textwidth,angle=270]{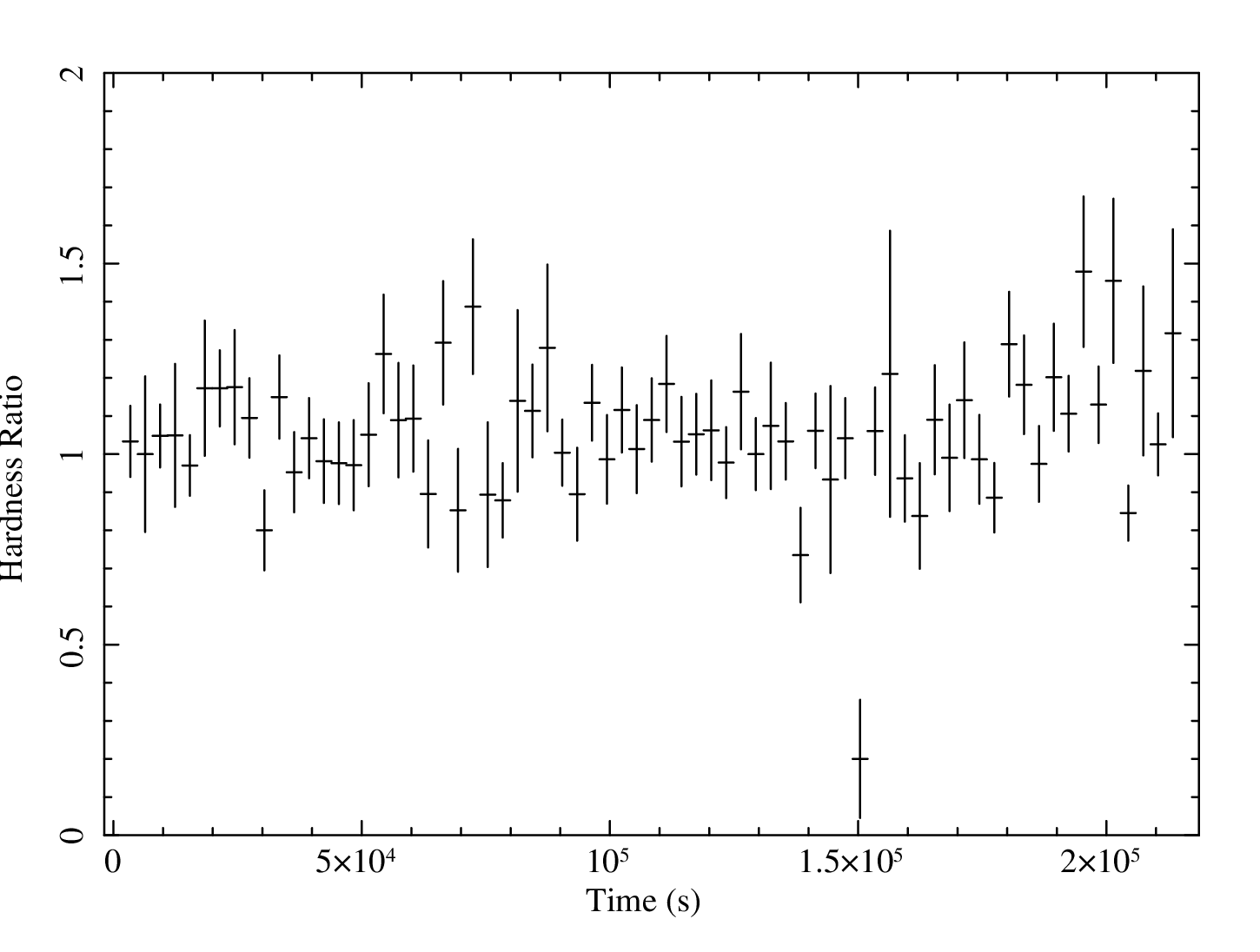}
\includegraphics[width=0.3\textwidth,angle=270]{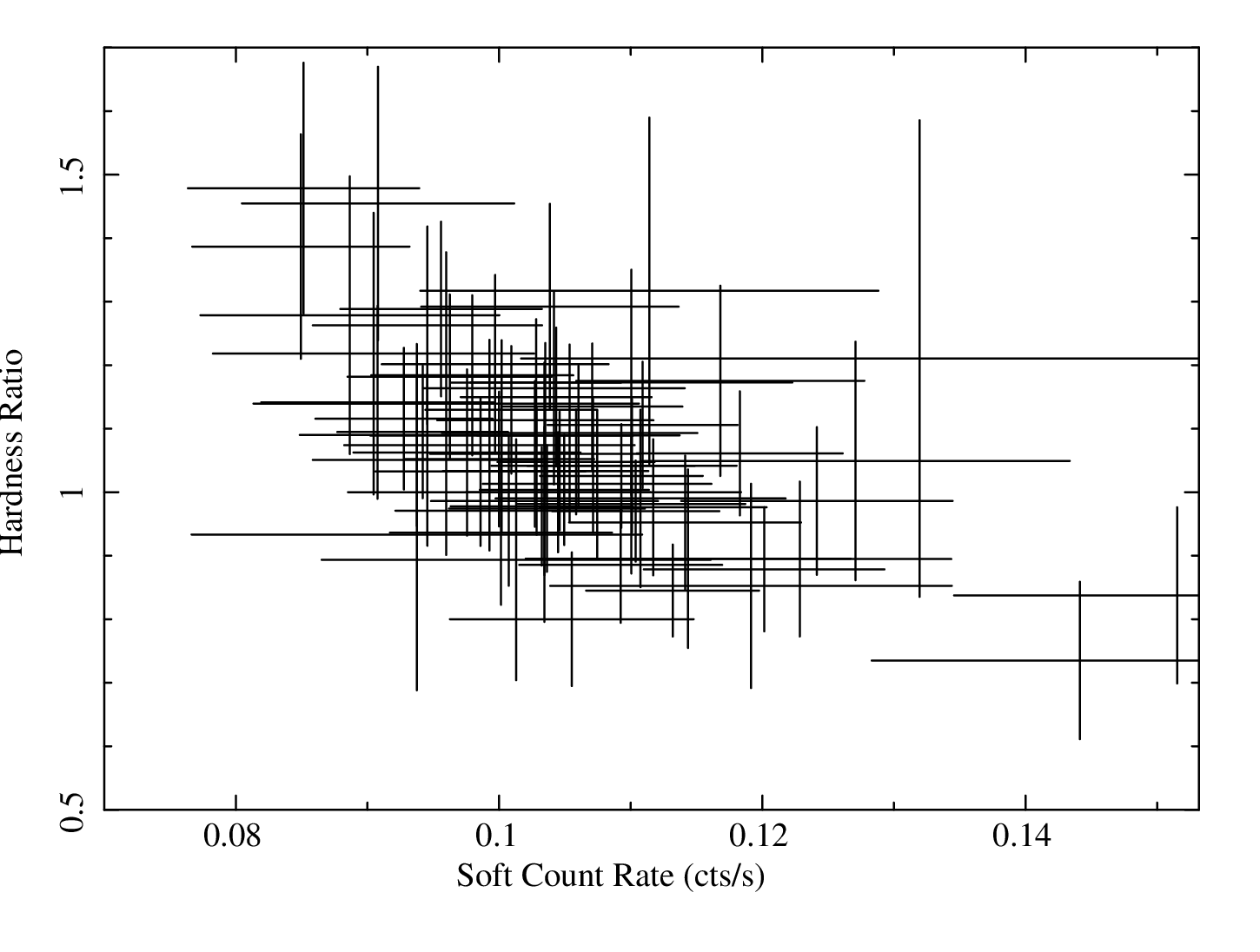}
\end{center}
\caption{\small {\it Left:} Hardness ratio plotted against time; {\it
  Right:} hardness ratio plotted against the soft flux over the
  course of the 2007 {\it Suzaku} observation of NGC~1052.  Though
  there is no significant change in hardness ratio vs. time during the
  observation, note that as the soft flux drops slightly, the source
  becomes harder.}
\label{fig:hr}
\end{figure}

No significant changes were seen in the count rates of any of the XIS
instruments or the PIN instrument over the course of the observation.
Fig.~\ref{fig:lcurves} depicts the light curves from XIS~0, XIS~1 and
XIS~3, as well as the combined XIS light curve vs. that of the PIN.
While small variations in count rate (over $\sim 10^4 \s$ timescales) were seen in both
data sets, neither varied more than a factor of two from its baseline
flux throughout the observation.  This flux was 
roughly $0.07 \ctsps$ for the PIN and $0.38 \ctsps$ for the combined XIS
data ($0.16$ for XIS~0, $0.14$ for XIS~3 and $0.21$ for XIS~1).
Fitting the combined XIS light curve to a constant model resulted in
$\chi^2/\nu=227/236\,(0.96)$ (where $\nu$ denotes the number of
degrees of freedom of the model: $\nu=\#$ detector channels -
  $\#$ free model parameters), and an uncertainty of $<0.01 \ctsps$ on
the $0.38 \ctsps$ count rate.  The PIN light curve fit to a constant
was not as good a fit at $\chi^2/\nu=351/245\,(1.43)$, with the
uncertainty in its $0.07 \ctsps$ count rate at $<0.005 \ctsps$.  We also
computed the hardness ratio vs. time for the combined XIS data, which is
consistent with unity as one might expect for a source without
significant variability over the course of the observation: $HR=1.02
\pm 0.02$, with $\chi^2/\nu=101/71\,(1.42)$.  Examining the
hardness ratio vs. soft band count rate, we note a significant
anti-correlation between the two: NGC~1052 becomes harder as the source flux
decreases.  This variation is best fit with a linear model: $HR=m{\rm
  (flux)}+b$, where $HR$ denotes the hardness ratio, ``flux'' is the soft
band count rate, $m$ is a multiplicative term and $b$ is a constant
term.  For the combined XIS data, $m=-8.35 \pm 2.31$ and $b=1.92 \pm
0.25$, with $\chi^2/\nu=66/71\,(0.93)$.  We have plotted the hardness
ratio vs. time and vs. soft count rate in Fig.~\ref{fig:hr}.

This hardening of the source
with decreased count rate demonstrates in a model-independent way that
the spectral variability of of NGC~1052 in X-rays is consistent with
the behavior observed in other Seyfert or Seyfert-like AGN (e.g.,
\citet{Papadakis2002,Markowitz2004}),
and contrasts with the behavior observed in jet-dominated AGN (e.g.,
\citet{Zhang2006}).  In other words, the anti-correlation of the hardness ratio with
the source flux confirms that the bulk of the X-ray emission in
NGC~1052 is not produced by the jet.
We have plotted the hardness ratio vs. time and vs. soft count rate in Fig.~\ref{fig:hr}.

The overall $2-10 \keV$ source flux
did not vary significantly, averaging to
$F_{2-10}= 5.37 \times 10^{-12} \ergpcmsqps$ (absorbed).  This was
split between the continuum and the Fe K lines, where the continuum
flux averaged $F_{\rm cont}= 4.86 \times 10^{-12} \ergpcmsqps$ and that of the combined broad and
narrow Fe K$\alpha$ lines averaged $F_{\rm Fe}= 5.10 \times 10^{-13}
\ergpcmsqps$.  The average unabsorbed $2-10 \keV$ flux at the distance of
NGC~1052, $F_{2-10}= 8.96 \times 10^{-12}\ergpcmsqps$, works out to an intrinsic luminosity of
$L_{\rm 2-10}= 4.60 \times 10^{41} \ergps$ using cosmological
parameters consistent with the WMAP5 $\Lambda$CDM-model results,
within errors: $H_0=70 \kmpspmpc$, $q_0=0.00$ and $\Lambda_0=0.73$.
These results, and the models used to derive them, are presented in more detail in \S4.

\section{Spectral Analysis}
\label{sec:analysis}

We began our analysis of the 2007 {\it Suzaku} spectrum of NGC~1052 by
examining the data from the operational XIS detectors.
We use the good signal-to-noise $0.5-9.0 \keV$ data to extract
information about the underlying continuum,
any soft excess emission, the Fe K line complex and any intrinsic
absorption in the system.  NGC~1052 is thought to exhibit
evidence for an AGN disk (the broad Fe K$\alpha$ line) and jets (radio
observations), as well as substantial, extended soft emission.
As such, we expected the spectrum to be a conglomerate of all of
these features since {\it Suzaku} lacks the spatial resolution
necessary to physically separate them.

We also analyzed the HXD/PIN spectrum from $12-35 \keV$ in order to
learn more about the nature of the continuum and to constrain the
amount of reflection seen in the source.  Because a broad iron line
component has
been observed in NGC~1052 in the past \citep{Guainazzi2000}, we expected that residual
emission above a power-law should remain at high energies due to the
presence of the so-called ``Compton hump.'' This feature has been
noted in many AGN with broad iron lines (for a review, see \citet{R+N2003}),
representing Compton down-scattering of the hard X-ray photons by the
material in the disk.  This Compton hump typically peaks at $\sim 20-30 \keV$, and is
thought to go hand-in-hand with the presence of a broad iron line, as
both spectral features arise from the same physical process of reflection.

\subsection{XIS Continuum}
\label{sec:xis}

\begin{figure}
\begin{center}
\includegraphics[width=0.6\textwidth,angle=270]{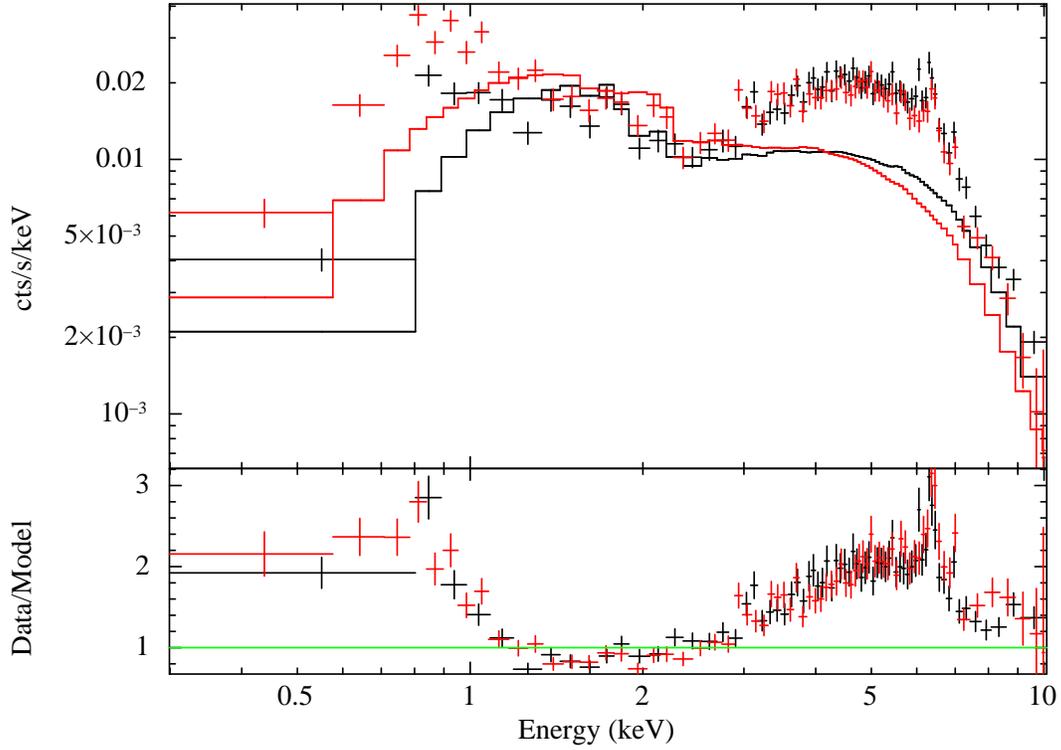}
\end{center}
\caption{\small The XIS spectrum of NGC~1052 fit with a power-law modified by
Galactic photoabsorption over the $0.5-3$ and $7-9 \keV$ bands
uninfluenced by the fluorescent Fe K$\alpha$ line.  The fit is
poor: $\chi^2/\nu=840/361\,(2.33)$.  Note especially the large
residuals at both the hard and soft ends.  Black crosses represent the
combined XIS~0+3 spectrum while the black line represents the model.
Red crosses and line represent the XIS~1 data and model.  The lower
panel of the figure shows the data-to-model residuals, with the green
line representing a perfect fit with a data-to-model ratio of one.}
\label{fig:phabs_po}
\end{figure}

Before the iron line region could be considered, it was necessary to first accurately
model the underlying continuum of NGC~1052.  We began by ignoring the
energy range where the iron line is thought to be important ($3-7
\keV$) and modeled the $0.5-3$ and $7-9 \keV$ energy bands with a
simple power-law modified by Galactic photoabsorption ($N_{\rm H}=2.83
\times 10^{20} \pcmsq$, as per \citet{Kalberla2005}).  The result was a
poor fit: $\chi^2/\nu=840/361\,(2.33)$.  Large residuals
remained at both high and low energies (see Fig.~\ref{fig:phabs_po}), and
a very hard photon index
for the power-law component was seen: $\Gamma=0.34 \pm 0.12$.  To
account for differences in calibration between the XIS~0+3 and the XIS~1
detectors, we multiplied each data set by a constant 
cross-normalization factor.  This constant was held at $1.00$ for XIS~0+3, and achieved a
best-fit value of $0.94 \pm 0.01$ for the XIS~1 data.

An apparent soft excess was clearly visible below
$\sim 1.5 \keV$ (see Fig.~\ref{fig:phabs_po}).  To mitigate this residual feature we incorporated a
thermal {\tt mekal} component \citep{Mewe1985}.  The fit
improved dramatically to $\chi^2/\nu=381/358\,(1.06)$
($\Delta\chi^2/\Delta\nu=-459/-3$, or an improvement in the global
goodness-of-fit by $459$ chi-squared points for an additional $3$
degrees of freedom), with
$kT=0.6 \pm 0.05 \keV$ and a solar abundance for the
{\tt mekal} component.  This component has been noted in previous
X-ray analyses of NGC~1052 by \citet{Kadler2004a} with {\it Chandra}
and \citet{Weaver1999} with {\it ASCA} (hereafter W99), and the temperature and flux
obtained from our {\it Suzaku} fit are comparable to both values in
both previous works.  This thermal emission most likely originates in
extended emission around the nucleus from the interaction of
the jets with the ISM \citep{Kadler2004a}.
Residuals still remained at low energies,
however, so we added in a second {\tt mekal} component (with an
abundance equal to that of the first) to try to
account for these features.  Although marginal improvement in the overall
goodness-of-fit was seen ($\Delta\chi^2/\Delta\nu=-12/-2$), the
addition of this second thermal component rendered the parameters of both {\tt mekal}
components unconstrained in an error analysis.  We therefore
removed the second thermal component.  Though modest residuals
remained at select energies,
no statistically significant improvement in fit
was achieved with the addition of discrete Gaussian components, as one
might expect in a photoionized plasma, or with allowing the abundance
of the {\tt mekal} component(s) to vary.
Alternatively, we also tried to account for this excess soft
emission with a second power-law, with and without further
absorption.  Both models exhibited significantly poorer fits than
that of the single {\tt mekal} component. 

Because NGC~1052 is also classified as a Seyfert 2 galaxy, it is reasonable to assume that
some intrinsic absorption exists in or around the nuclear region.
Such absorption could play a significant role in the spectral
curvature seen, so it must be properly accounted for in
the continuum model.  We employed a partial-covering absorber ({\tt
  zpcfabs}) intrinsic to the source to model this component, applying it only to the power-law
and other components of nuclear AGN emission and not to the extended
{\tt mekal} components.  A statistically significant improvement in
fit was seen: $\chi^2/\nu=335/356\,(0.94)$, with $N_{\rm H}=1.08
\pm 0.15
\times 10^{23} \pcmsq$ and a covering fraction of $84 \pm 2\%$.  This
marked an improvement in fit of $\Delta\chi^2/\Delta\nu=-46/-2$.
With the inclusion
of this intrinsic absorption, the
power-law photon index increased to $\Gamma=0.80 \pm 0.09$.  This was a
significant increase in photon index from the fit without intrinsic
absorption. 

Physically, this continuum model and its parameters were roughly comparable to
those which were fit for the 1996
{\it ASCA} observation of the source (W99) and the 2000 {\it BeppoSAX}
observation by \citet{Guainazzi2000} --- hereafter G00 --- although the
photon index of the power-law is considerably harder and the intrinsic
absorption less dense in this 2007 observation.  A model with a dual neutral
absorber modifying intrinsic plus scattered power-law components as in
W99 and G00 did not improve the fit, statistically; moreover,
the parameter values of the second power-law and absorber could not be
constrained.

Even after modeling the neutral absorption in the system, some
residual spectral curvature still remained around $2-3
\keV$, indicating that some further component of absorption
likely remained unmodeled.  We applied the simple ionized absorber
model {\tt absori} \citep{Zdziarski1995} to the nuclear emission and noted a
marked visual improvement to the fit from $2-3 \keV$, as well as a
small improvement in the global goodness-of-fit to
$\chi^2/\nu=325/354\,(0.92)$, an improvement of
$\Delta\chi^2/\Delta\nu=-10/-2$.  This absorber was reasonably
thick and moderately ionized, with a column density of $N_{\rm H}=1.37
\pm 0.02
\times 10^{22} \pcmsq$
and an ionization parameter of $\xi=68 \pm 7 \ergpcmps$.  With the addition
of this component, the power-law photon index increased to
$\Gamma=1.50 \pm 0.02$.
This value lies within reasonable, expected physical limits for the X-ray
continuum of an AGN.  Our base continuum model therefore included a
power-law modified by intrinsic absorption from both a cold, patchy
absorber and a moderately ionized absorber of smaller column density,
along with extended thermal emission.  All model components were
affected by Galactic photoabsorption.

\subsection{The Fe K$\alpha$ Line}
\label{sec:fe}

\begin{figure}
\begin{center}
\includegraphics[width=0.3\textwidth,angle=270]{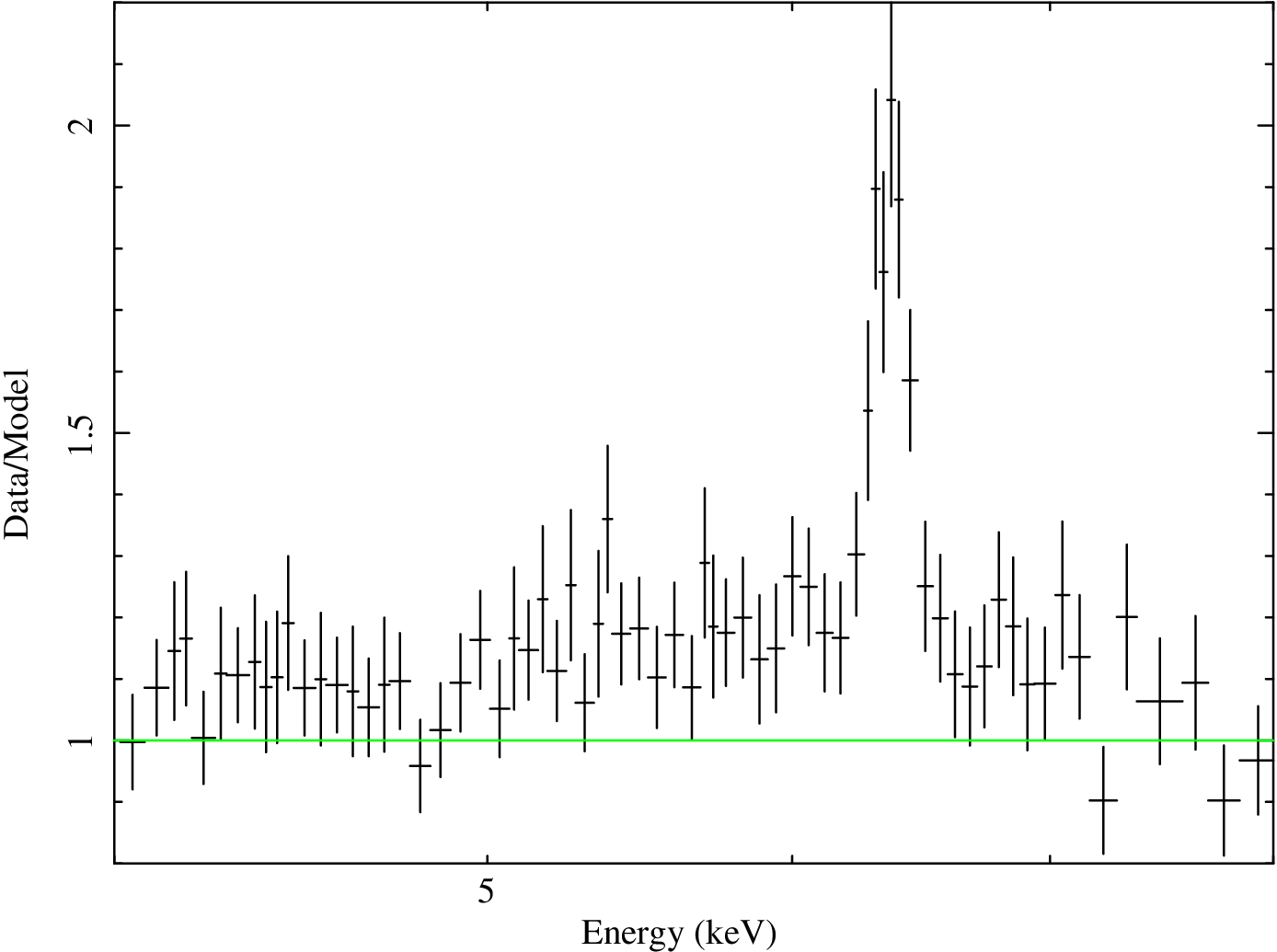}
\end{center}
\begin{center}
\includegraphics[width=0.3\textwidth,angle=270]{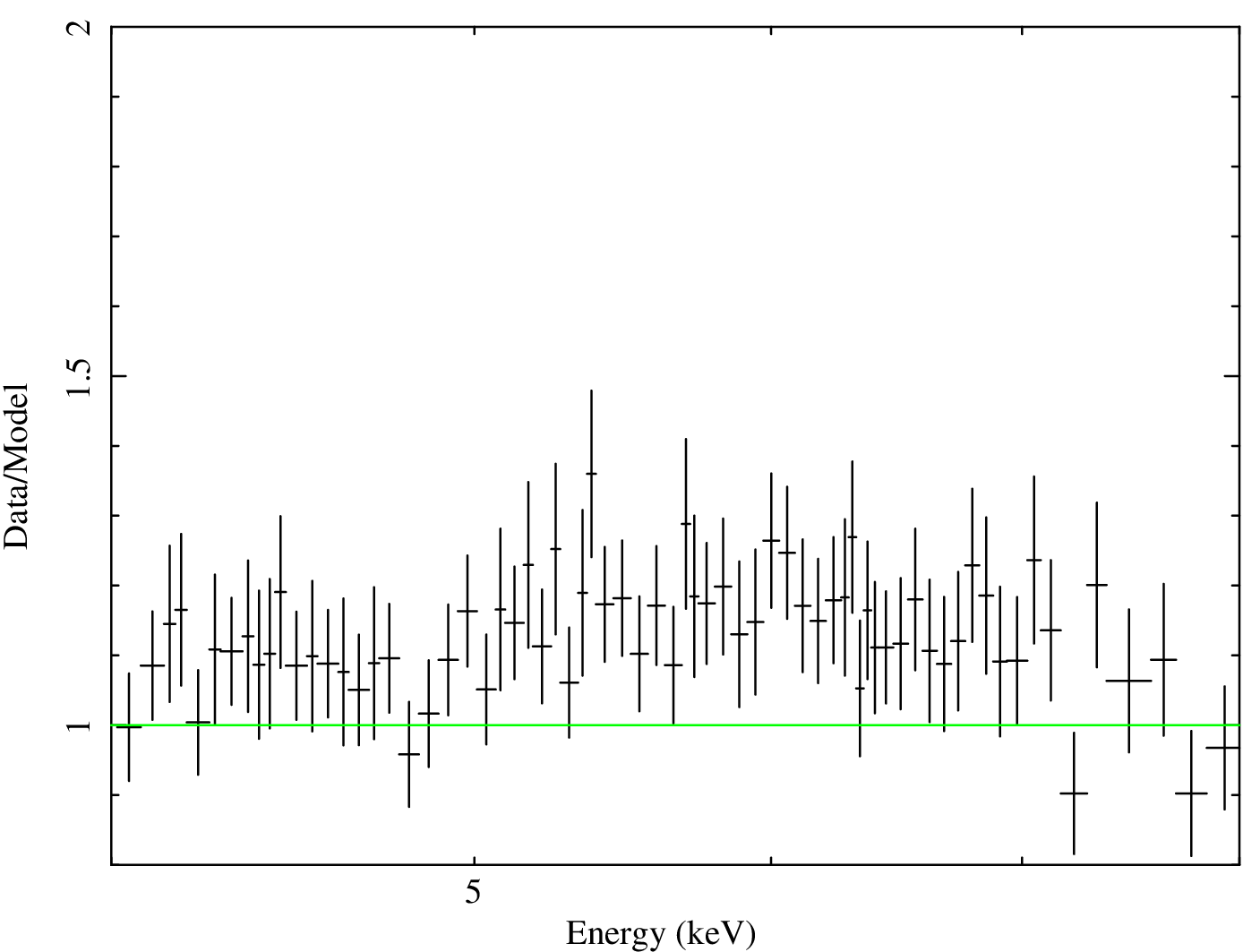}
\end{center}
\begin{center}
\includegraphics[width=0.3\textwidth,angle=270]{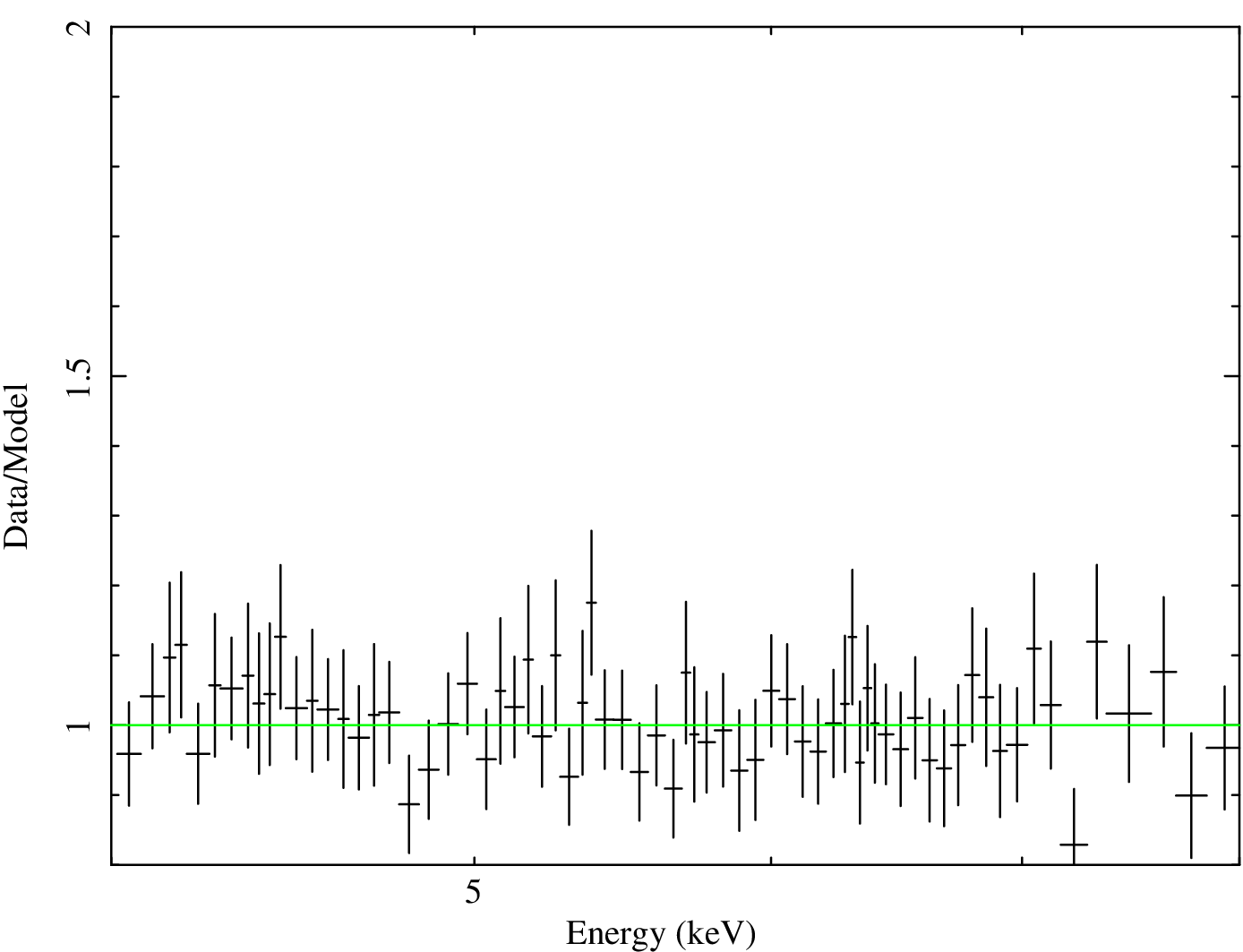}
\end{center}
\caption{\small {\it Top:} The residual emission feature remaining after fitting the
  continuum of the NGC~1052 {\it Suzaku}/XIS data, as per \S\ref{sec:xis}.  Note the
  prominent peak centered at $6.4 \keV$ in the rest frame and the excess
  emission extending down to $\sim 4 \keV$.
  $\chi^2/\nu=711/631\,(1.13)$.  {\it Middle:} Residuals remaining after
  fitting only the narrow Fe K$\alpha$ core.  Evidence for a broader
  feature is clear.  $\chi^2/\nu=584/629\,(0.93)$.  {\it
  Bottom:} Residuals remaining after a {\tt laor} line
  \citep{Laor1991} was used to
  model the broad Fe K$\alpha$ emission line as well.  $\chi^2/\nu=561/625\,(0.90)$.}
\label{fig:fek}
\end{figure}

Having fit the continuum, we then took the
energies from $3-7 \keV$ back into consideration and attempted to model
the prominent residuals remaining, highlighted by a strong emission
feature centered at $6.4 \keV$ in the rest frame that was assumed to be
neutral Fe K$\alpha$ (see Fig.~\ref{fig:fek}, top).  We began by holding
the continuum parameters constant, except for normalizations and the
power-law photon index, and refitting.  The strong residual at $6.4 \keV$
remained, appearing to have a broadened red wing associated with it
that extended down to $\sim 4 \keV$.  For this fit, $\chi^2/\nu=711/631\,(1.13)$.  

To account for the strong emission feature of $6.4 \keV$, we initially
assumed some core contribution from neutral Fe K$\alpha$ outside of the inner
accretion disk in distant material.  This would produce a relatively
narrow emission line, being
outside the region where relativistic broadening becomes important.
Inserting such a line ($\sigma=5 \eV$, intentionally less than the
resolution of the XIS in order to make the line truly narrow)
into the model yielded a vast improvement in fit to
$\chi^2/\nu=584/629\,(0.93)$ for $\Delta\chi^2/\Delta\nu=-127/-3$ over
the case with no line present.  The energy of the line was constrained
in the fit to $E=6.39-6.41 \keV$,
indicative of neutral iron, and the feature had an equivalent width
$EW=88-134\eV$.  These parameter
ranges are quoted at $90\%$
confidence for one interesting parameter; this confidence interval
applies to all parameter ranges quoted in this Section.
The broad residuals centered on $6.4 \keV$ still
remained, however (see Fig.~\ref{fig:fek}, middle).  

We attempted to model these broad residuals three separate ways: with a
broad Gaussian, with a {\tt diskline} component
representing emission from the inner disk around a Schwarzschild black
hole \citep{Fabian1989}, and with a {\tt laor} component representing
emission from the inner disk around a maximally-spinning Kerr black
hole \citep{Laor1991}.  Each line was centered at $6.4 \keV$ in the
rest frame, corresponding to neutral Fe K$\alpha$.  The Gaussian
component improved
the fit by $\Delta\chi^2/\Delta\nu=-13/-2$ to
$\chi^2/\nu=571/627\,(0.91)$, and was quite broad:
$\sigma=642-852 \eV$.  Its equivalent width of $EW_{\rm broad}=166-236
\eV$ was
slightly smaller than that of previous observations, e.g.,
W99 and G00, in which $EW_{\rm broad} \sim 300 \eV$.   
Relaxing the energy constraint of the line resulted in $E=5.41-5.88
\keV$, with $\sigma=780-1300 \eV$ and $EW_{\rm broad}=359-390 \eV$, though
significant residuals still remained on the red wing of the line.
We then turned to {\tt diskline} and {\tt laor} in
the hope of modeling the underlying broad feature, recognizing that
these models assume that the broad emission originates in the inner
accretion disk, which may not be the case if NGC~1052 harbors an ADAF.

The {\tt diskline} model yielded $\chi^2/\nu=574/625\,(0.92)$,
which is not significantly different from the Gaussian fit ($\Delta\chi^2/\Delta\nu=+3/-2$),
statistically, though it did account
for more of the residuals below $6 \keV$.  VLBI observations of
the radio jet on sub-parsec scales have constrained
the inclination angle of the accretion disk to be between
$57-72\degmark$ relative to our line of sight \citep{Kadler2004a}, so we applied
these constraints to the model.  Our best fit yielded an accretion disk emissivity
index of $\alpha=1.69-3.69$ (where the disk emissivity $\epsilon
\propto r^{-\alpha}$) and an inner disk radius of $r_{\rm
  in}<26\,r_{\rm g}$ (where $r_{\rm g}=GM/c^2$), but the inclination angle could not be further
constrained.  We held the outer radius constant at $r_{\rm out}=400\,r_{\rm g}$,
since it was unable to be constrained and our emissivity index
dictated that this radius safely encompasses all of the relevant
emission from the disk.  Fixing the inclination angle to an average
value of $65\degmark$ and the emissivity index of the disk to a
``typical'' Seyfert AGN value of $\alpha=3$ yielded no significant
improvement in fit or change in the other model parameter values, and
only resulted in a constraint on the inner radius of disk emission of
$r_{\rm in}=11-52\,r_{\rm g}$. 

The {\tt laor} model provided the best statistical fit at
  $\chi^2/\nu=561/625\,(0.90)$, an improvement of $\Delta\chi^2/\Delta\nu=-13/0$
  over the {\tt diskline} model and $\Delta\chi^2/\Delta\nu=-10/-2$ over the
  broad Gaussian line model.  The {\tt laor} model also appeared
  to be most effective at reducing
  the residuals between $4-7 \keV$ (see Fig.~\ref{fig:fek}, bottom).  Keeping the same
  disk inclination constraints mentioned above for the {\tt diskline}
  model, we again could not further constrain this parameter, but we noted
  that the emissivity index of the disk was fairly centrally concentrated at
  $\alpha=1.53-5.10$, while the inner radius of emission is only mildly
  constrained within $r_{\rm in}=8.93-45.10\,r_{\rm g}$.  The equivalent
  width of the broad iron line here was found to lie within $EW_{\rm broad}=110-297 \eV$.
  Interestingly, when
  we relaxed the constraint on the inclination angle of the disk we
  obtained a best-fit value of $i=45 \pm 5\degmark$; this is roughly a
  $\sim 2\sigma$ deviation from the radio constraints.  Fixing the
  inclination angle and emissivity index as above for the {\tt
  diskline} case worsened the global goodness-of-fit by 
$22$ chi-squared points for $2$ fewer degrees of freedom
  ($\Delta\chi^2/\Delta\nu=+22/+2$) and
no significant improvement on the constraint for the inner radius of disk emission:
$r_{\rm in}=13-44\,r_{\rm g}$.

\begin{figure}
\begin{center}
\includegraphics[width=0.6\textwidth,angle=270]{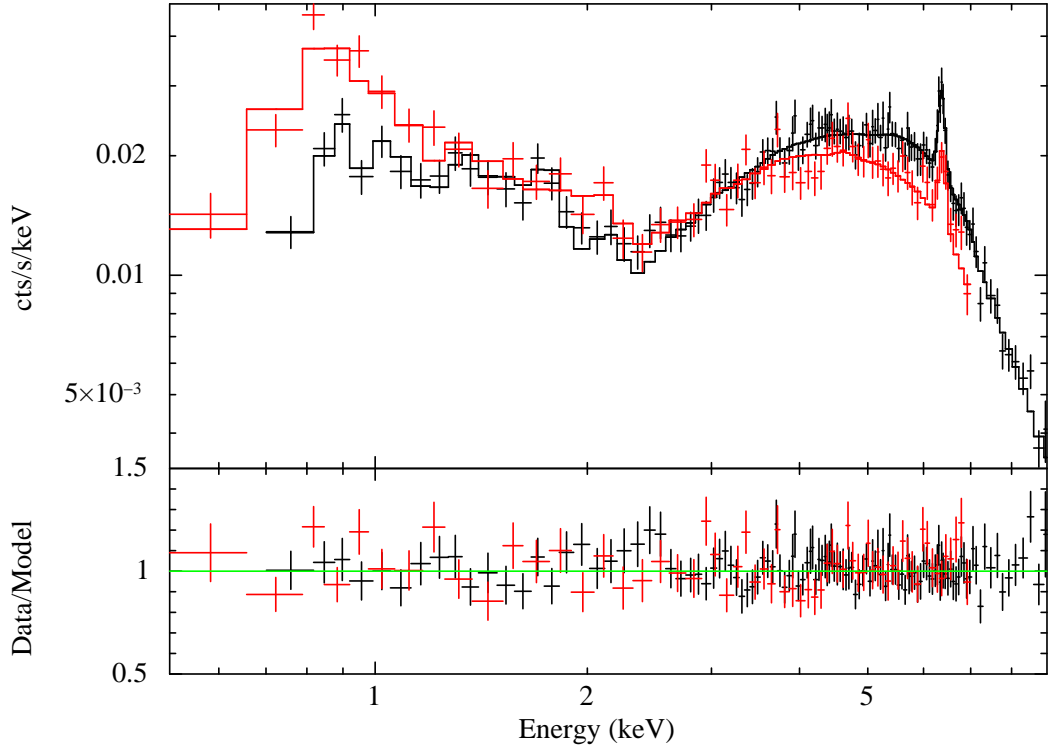}
\end{center}
\caption{\small The best-fitting model for the XIS data.   This model includes a
  power-law continuum modified by intrinsic absorption from a
  partial-covering neutral material as well as a mildly ionized gas, extended
  thermal emission, and both broad and narrow components of neutral
  Fe K$\alpha$.  The spectrum is also modified by Galactic
  photoabsorption.  $\chi^2/\nu=561/625\,(0.90)$. The red, black and
  green colors/lines are the same as in Fig.~1.}
\label{fig:laor}
\end{figure}

  The {\tt laor} fit was a slight improvement over both
  the Gaussian and {\tt diskline} models, though it should be
  noted that the uncertainty of the $r_{\rm in}$ parameter
  effectively renders it impossible to meaningfully distinguish between the {\tt
  diskline} and {\tt laor} models based on the Fe K$\alpha$ profile
  alone.  If the {\tt laor} model fit could better constrain the inner
  disk radius to $r_{\rm in}<6\,r_{\rm g}$ (indicating that an
  optically-thick disk extends down to small radii), it would suggest that the black
  hole at the center of NGC~1052 is rapidly rotating, as many theories 
  expect for a source that powers relativistic jets (e.g., \citet{BZ1977,Meier2001,Hawley2006}).  We cannot make
  this statement unequivocally based on our data, however.      
The data fit with this {\tt laor} model
are shown in Fig.~\ref{fig:laor}.  The final best-fit model
parameters and their errors, incorporating the HXD/PIN data as well, 
are presented in the following Section within Table~\ref{tab:tab1}
as Model~1.  

Absorption has often been suggested as an alternative to broad iron
emission when attempting to model enhanced spectral curvature in the
$4-7 \keV$ range (e.g., \citet{Kinkhabwala2003}).  To address this
possibility, we compared our best {\tt laor} fit above to a model in
which we removed the broad iron line and replaced it with a second
absorbed power-law component.  The resulting fit yielded
$\chi^2/\nu=582/627\,(0.93)$, as opposed to the
$\chi^2/\nu=561/625\,(0.90)$ achieved with the final {\tt laor} model
($\Delta\chi^2/\Delta\nu=+22/+2$), and left
significant residuals around $6.4 \keV$ and below.  Moreover, the
parameter values for both the partial-covering absorber and the
power-law components are not constrained in the fit.  The models
employing broad emission lines unquestionably provide
better statistical and visual fits to the data.

\subsection{Addition of the HXD/PIN Data}
\label{sec:pin}

Adding the HXD/PIN data on to the combined XIS~0+3 and XIS~1 spectra, we
noticed that the signal-to-noise ratio deteriorated significantly above
$\sim 35 \keV$ and below $\sim 12 \keV$.  As such, we restricted our energy range for this
detector to $12-35 \keV$, applying a constant cross-normalization factor as
we did with the XIS~1 data set.  In this case, we used a parameter value
of $1.18$ in keeping with the value expected
for the HXD nominal pointing position.

\begin{figure}
\begin{center}
\includegraphics[width=0.6\textwidth,angle=270]{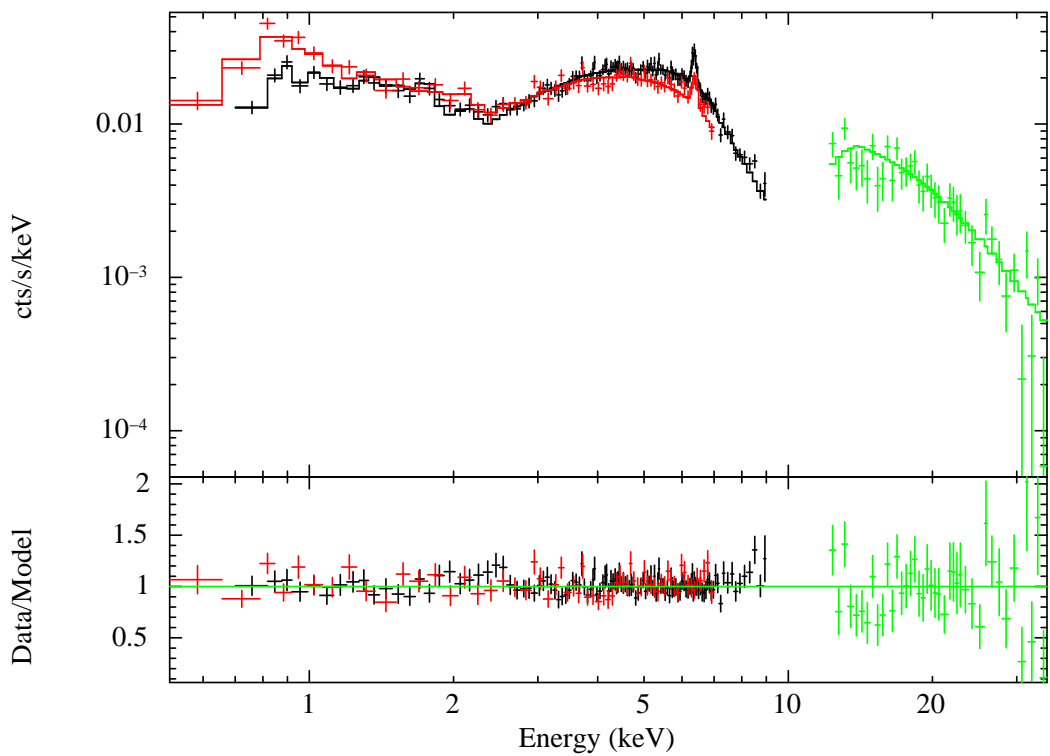}
\end{center}
\caption{\small HXD/PIN data included from $12-35 \keV$ and refitted with
  Model~1, which does not include any type of continuum disk
  reflection.  Inclusion of this data does not significantly alter the model
  parameter values from the $0.5-9 \keV$ fit.  Though a broad line is
  still required at $6.4 \keV$, no accompanying evidence of reflection
  from the disk above $9 \keV$ is seen.  Red and black colors follow
  the designations of Fig.~1.  In the top panel, the green crosses and line
  represent the PIN data and model, respectively.  In the bottom panel
  the horizontal green line still represents a data-to-model ratio of
  one, while the green crosses show the residuals between the PIN data
  and model.  $\chi^2/\nu=660/685\,(0.96)$.}
\label{fig:pin_resid_laor}
\end{figure}

We considered the combined data refitted with Model~1 (photoabsorbed
power-law, a partly-ionized absorber and a partial-covering neutral
absorber, soft thermal emission, narrow and broad Fe K$\alpha$ components), including
no reflection component above $9 \keV$.  Fig.~\ref{fig:pin_resid_laor}
shows the data, model and residuals from this fit, which yields
$\chi^2/\nu=660/685\,(0.96)$.  The inclusion of the high-energy
data left the $2-9 \keV$ parameters
effectively unchanged from their values in the fit without the PIN data.
If a significant flux contribution from reflection is present, as one might
assume if the broad iron line has an origin in a relativistic
accretion disk, we would expect
to see some excess in the emission over $9 \keV$ 
relative to Model~1.  No such feature was detected,
however.

Table~\ref{tab:tab1} shows the full $0.5-35 \keV$ fit with $90\%$
confidence errors.  

\subsection{Alternative Continuum Models}
\label{sec:alt}

Though no evidence for Comptonized reflection is apparent to the eye above
$9 \keV$, it is nonetheless important to try to place upper limits on
its possible contribution to the overall spectrum.  With this in mind,
we utilized two popular, well-regarded models of reflection from an
accretion disk around a black hole in an attempt to quantify the
amount of reflection that might be present in the spectrum of
NGC~1052.  

The {\tt pexrav} (or {\tt pexriv}) model of \citet{Magdziarz1995} is one of the most
widely used public models for AGN emission which incorporates reflected continuum
from a neutral (or partly ionized) accretion disk.  While it represents a more physical model
than the simple power-law alone, it does not include line emission
from the disk (e.g., Fe K$\alpha$) and it is difficult, in practice,
to constrain the three most important parameters simultaneously:
photon index, reflection fraction and disk inclination angle.  

Adding a {\tt pexrav} component to our power-law
continuum (and keeping the inclination angle of the disk constant at
the {\tt laor} value, the power-law cutoff at $100 \keV$, and the
abundances fixed at their solar values), we found
$\chi^2/\nu=660/683\,(0.97)$, only a very marginal change in fit from that
of Model~1 over the $0.5-35 \keV$ energy range
($\Delta\chi^2/\Delta\nu=0/-2$).  The photon index fit between
$\Gamma=1.45-1.56$ while the reflection fraction was constrained to be
quite small: $R_{\rm refl}<0.01$.  As we expect, based on the residuals from
fitting Model~1, reflection is not a significant contributor to the
overall observed spectrum.  Though a {\tt pexriv} component was also
fit, the ionization parameter was unable to be constrained, so we
elected to assume a neutral disk.  In this fit,
the emissivity index of the {\tt laor} line for
Fe K$\alpha$ had very similar constraints to those found in Model~1:
$\alpha=1.74-4.56$.  Also, to $90\%$ confidence, the inner radius of
emission in the disk was constrained to $r_{\rm in} \leq
45\,r_{\rm g}$, as we found in Model~1.  The equivalent width of the
broad Fe K$\alpha$ line in this model
also remained consistent with that found in Model~1, at $EW=113-293 \eV$.
The {\tt pexrav} model will hereafter be known as Model~2.  Its final
parameter values and their errors are also presented in Table~\ref{tab:tab1}.

The {\tt reflion} model of \citet{Ross2005} is another widely
available code used to describe the reflected spectrum of an accretion
disk around a black hole.  The advantage of this model over the {\tt
  pexrav} model is that it is self-consistent: fluorescent emission lines from
many ionized species are
included in addition to the continuum, most notably those of the Fe K
complex.  The ionization
parameter, incident power-law photon index and iron abundance are
specified by the user.  The disadvantage is that photon indices below
$\Gamma=1$ are outside the allowed parameter range of the model, and
there is not a parameter for the reflection fraction, as in {\tt
  pexrav}.  The user is therefore obliged to estimate it using, e.g., 
\begin{equation}
R_{\rm refl} = \frac{K_{\rm refl}}{K_{\rm po}}\,
\frac{F_{\rm refl}}{F_{\rm po}}\, \left(\frac{\xi}{30}\right)^{-1}.
\end{equation}
Here $K_{\rm refl}$ and $K_{\rm po}$ denote the normalizations of the
{\tt reflion} and power-law components from our best-fit spectral
model, $\xi$ is the best-fit ionization parameter characterizing the
disk reflection (where $\xi=30$ denotes a neutral disk in the model), and $F_{\rm
  refl}$ and $F_{\rm po}$ are the total
fluxes contained in the {\tt reflion} and power-law components for
unity normalization over the full wavelength range ($0.001-1000
\keV$).  

We began by removing the {\tt
  laor} line from the Model~1 fit and adding in a {\tt reflion}
component, with the iron abundance held at the solar value and the
photon index linked to that of the power-law.  As expected, this fit
mimicked that of the continuum plus a narrow $6.4 \keV$ Gaussian,
leaving residuals on the low-energy end of the line.  Statistically,
$\chi^2/\nu=679/687\,(0.99)$.  Allowing the iron abundance to fit freely
yielded a small improvement in fit, with $\chi^2/\nu=667/686\,(0.97)$
  for ${\rm Fe/solar} \leq 3.53$, but doing so rendered
the ionization parameter and normalization unconstrained.  As such we
  elected to keep the iron abundance constant at the solar value.

\begin{figure}
\begin{center}
\includegraphics[width=0.3\textwidth,angle=270]{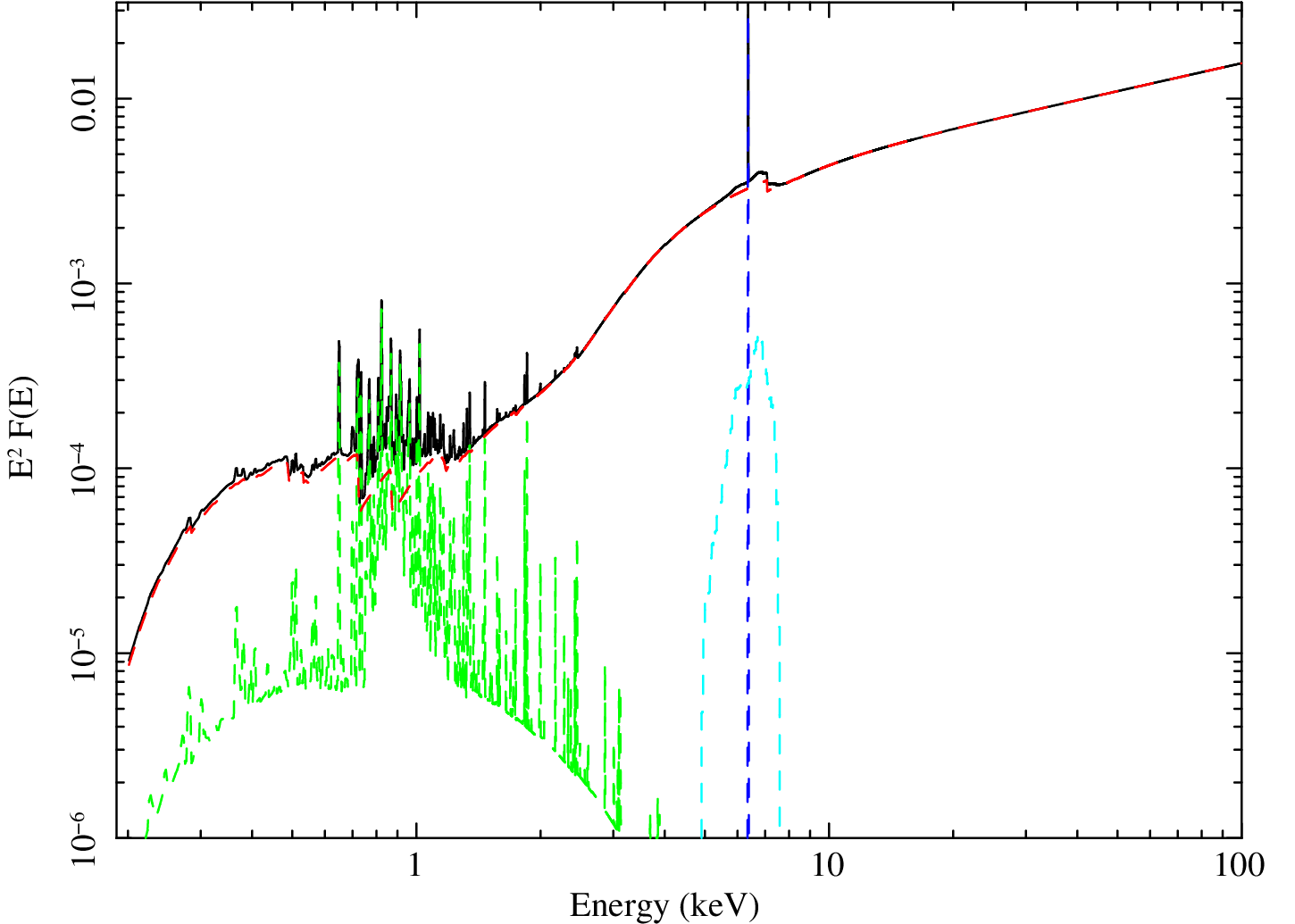}
\end{center}
\begin{center}
\includegraphics[width=0.3\textwidth,angle=270]{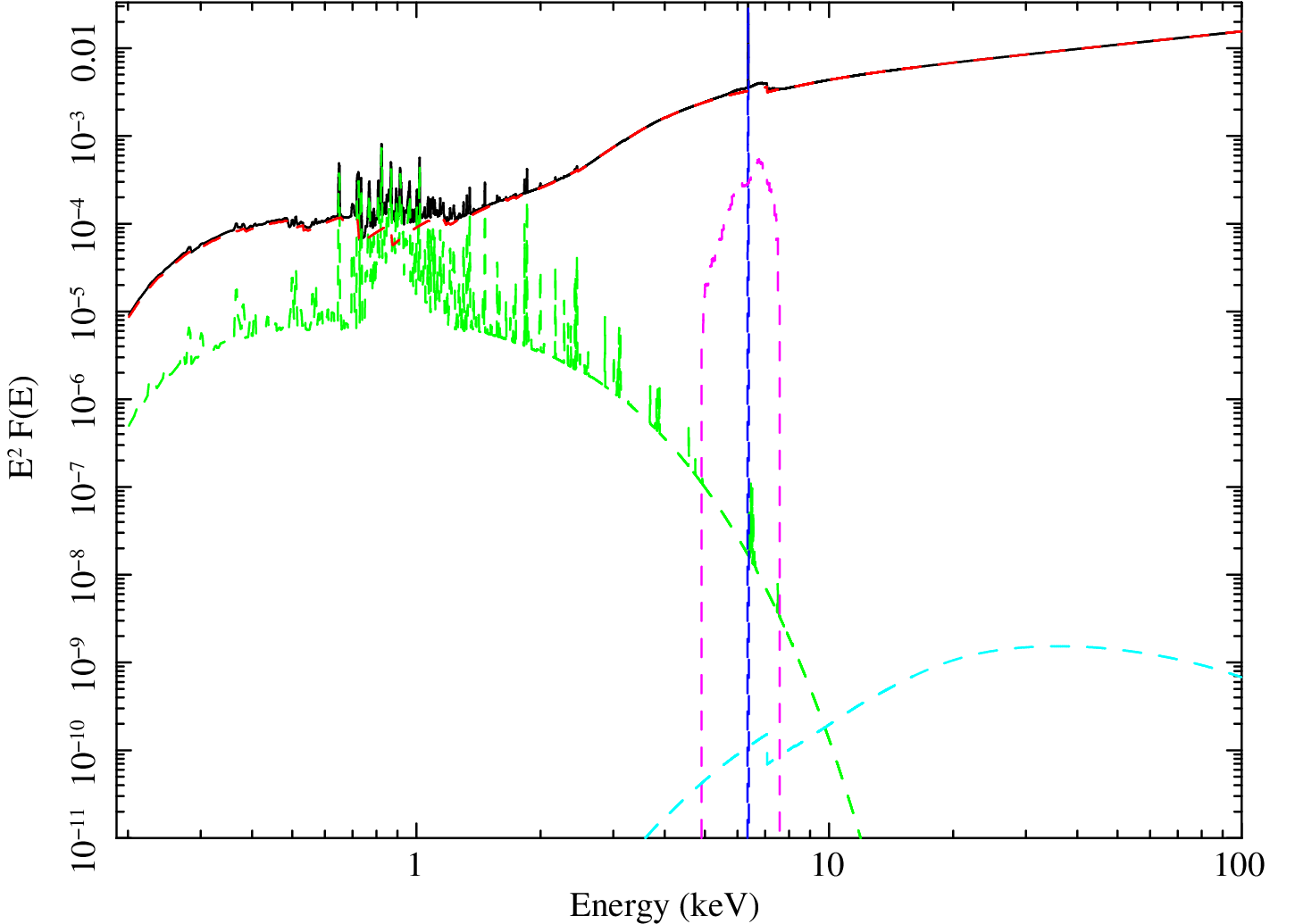}
\end{center}
\begin{center}
\includegraphics[width=0.3\textwidth,angle=270]{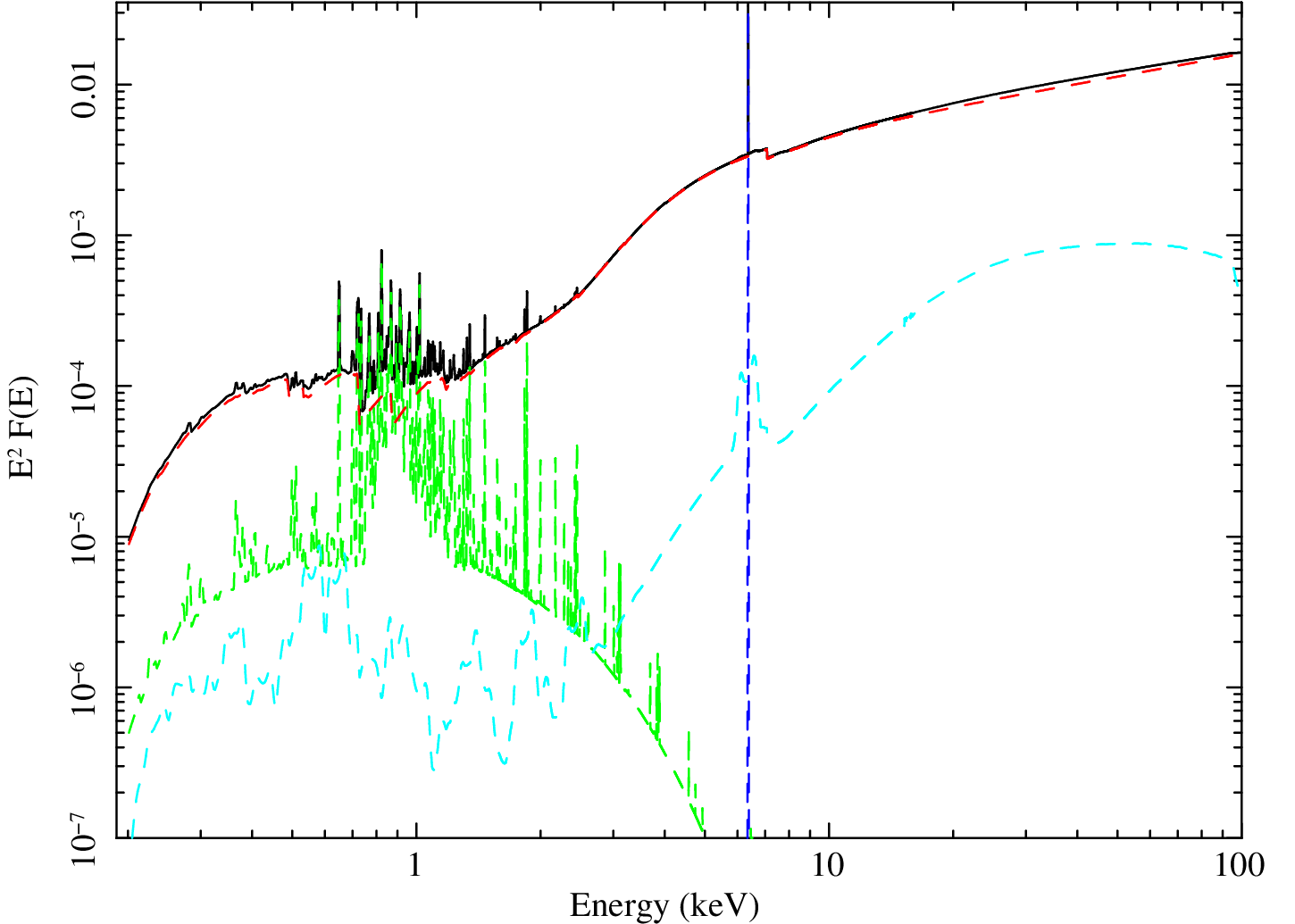}
\end{center}
\caption{\small The relative contributions of each
  model component in the three relevant fits to the NGC~1052 XIS and
  PIN data:
  {\tt laor} (top, Model~1), {\tt pexrav+laor} (middle, Model~2) and {\tt kdblur(reflion)}
  (bottom, Model~3).  Each model presents its own strengths and weaknesses, as
  discussed in the text.  Models are extended to $100 \keV$ to
  illustrate their projected contributions at higher energies.  Red
  represents the power-law (affected by Galactic and intrinsic
  absorption), green represents the thermal {\tt mekal}
  component, Gaussian emission lines are in dark blue and reflection
  features such as the {\tt laor} line (top), the {\tt pexrav}
  component (middle) and the {\tt kdblur(reflion)} model (bottom) are
  in light blue.  The {\tt laor} line in Model~2 is in magenta.}
\label{fig:models}
\end{figure}

If NGC~1052 harbors a rapidly spinning black hole, as our best {\tt
  laor} fit seems to indicate, we must take into account the relativistic
  effects on emission originating from the inner accretion disk.  We
  incorporated this smearing of spectral features
  using the {\tt laor}-based convolution model {\tt kdblur}.  Our best
  fit was $\chi^2/\nu=660/684\,(0.96)$, comparable to our
  best {\tt laor} and {\tt pexrav} fits ($\Delta\chi^2/\Delta\nu=0/-1$ and
  $\Delta\chi^2/\Delta\nu=0/+1$, respectively).  However, we were unable to
  constrain the emissivity index of the accretion disk, though we did
  obtain constraints on the inner radius of emission that are
  consistent with those found in the {\tt laor} model: $r_{\rm
  in}<45\,r_{\rm g}$.  The estimated reflection fraction is
  quite low at $R_{\rm refl} \leq 0.006$, consistent with the results of the
  {\tt pexrav} model fit and partly attributable to the moderately high
  ionization parameter ($\xi \leq 111$) in the {\tt reflion} model.  
The {\tt kdblur(reflion)} model will hereafter be known as Model~3. 
For comparison, the model components for the Models~1-3 fits are plotted in
  Fig.~\ref{fig:models}, and their
  best-fit parameters and errors are listed in Table~\ref{tab:tab1}.  

Models~2-3 constrain the amount of reflection from the disk to
be $R_{\rm refl} \leq 0.01$, consistent with the lack of a discernible
Compton hump above $10 \keV$ in the HXD/PIN data.  Moreover, as we
describe in detail later in \S\ref{sec:discussion}, if the broad iron
line component results from fluorescence in the disk caused by incident hard
X-ray emission, then the equivalent width of the Fe K$\alpha$ line
should be directly related to the normalization of the reflection
component in our model as compared to that of the power-law.  Yet this
is not the case in our data.  Based on the low normalizations for
reflection seen in both Model~2 and Model~3, we should expect a broad
Fe K$\alpha$ equivalent width of $EW \sim 80 \eV$, much less than the
observed $EW \sim 185 \eV$ in both Models~1-2.  In other words, our
broad iron line does not appear to have a correspondingly strong
amount of continuum reflection associated with it.  It is possible
that this ``missing'' reflection is actually present to some greater
degree in the data, but is perhaps drowned out by the power-law
continuum.  It is also possible that reflection is simply absent in
this source.  This scenario presents several intriguing possibilities
for physical interpretations, which we discuss in
\S\ref{sec:discussion}. 

\begin{table}
\begin{center}
\begin{tabular}{|l|l|l|l|l|}
\hline \hline
{\bf Component} & {\bf Parameter Value} & {\bf Model~1} & {\bf Model~2} & {\bf Model~3}\\
\hline
{\tt mekal} & $kT\,(\keV)$ & $0.64^{+0.04}_{-0.04}$ & $0.64^{+0.04}_{-0.04}$ & $0.64^{+0.04}_{-0.04}$ \\
& $K_{\rm kT}\,(\phpcmsqps)$ & $4.14^{+1.00}_{-1.02} \times 10^{-5}$ &
$4.14^{+0.74}_{-0.84} \times 10^{-5}$ & $4.14^{+0.64}_{-0.57} \times
10^{-5}$ \\
& $F_{\rm kT}\,(\ergpcmsqps)$ & $9.00^{+2.17}_{-2.22} \times 10^{-14}$ & $9.48^{+1.69}_{-1.92} \times
10^{-14}$ & $8.78^{+1.35}_{-1.94} \times 10^{-14}$ \\
\hline
{\tt absori} & $\Gamma$ & $1.50^{+0.07}_{-0.08}$ & $1.50^{+0.06}_{-0.05}$ & $1.50^{+0.05}_{-0.05}$ \\
& $N_{\rm H}\,(\times 10^{22} \pcmsq)$ & $1.37^{+0.54}_{-0.78}$ & $1.37^{+0.27}_{-0.23}$ & $1.45^{+0.63}_{-0.53}$ \\
& $T\,(\K)$ & $3 \times 10^4$ & $3 \times 10^4$ & $3 \times 10^4$ \\
& $\xi\,(\ergpcmps)$ & $68^{+100}_{-31}$ & $66^{+36}_{-46}$ & $67^{+26}_{-33}$ \\
\hline
{\tt zpcfabs} & $N_{\rm H}\,(\times 10^{22} \pcmsq)$ & $10.82^{+0.44}_{-0.77}$ & $10.82^{+0.64}_{-0.94}$ & $10.83^{+0.55}_{-0.87}$ \\
& $\% {\rm cover}$ & $0.84^{+0.03}_{-0.04}$ & $0.84^{+0.02}_{-0.02}$ & $0.85^{+0.02}_{-0.02}$ \\ 
\hline
{\tt zpo} & $\Gamma$ & $1.50^{+0.07}_{-0.08}$ & $1.50^{+0.06}_{-0.05}$ & $1.50^{+0.05}_{-0.05}$ \\ 
& $K_{\Gamma}\,(\phpcmsqps)$ & $1.55^{+0.20}_{-0.24} \times 10^{-3}$ &
$1.55^{+0.21}_{-0.25} \times 10^{-3}$ & $1.62^{+0.23}_{-0.35} \times
10^{-3}$ \\
& $F_{\Gamma}\,(\ergpcmsqps)$ & $1.89^{+0.24}_{-0.29} \times 10^{-11}$ & $1.79^{+0.24}_{-0.29} \times
10^{-11}$ & $1.85^{+0.34}_{-0.32} \times 10^{-11}$ \\
\hline
{\tt zgauss} & $E\,(\keV)$ & $6.40^{+0.01}_{-0.01}$ & $6.40^{+0.01}_{-0.01}$ & $6.40^{+0.01}_{-0.01}$ \\
& $K_{\rm narrow}\,(\phpcmsqps)$ & $1.18^{+0.19}_{-0.19} \times 10^{-5}$
  & $1.19^{+0.19}_{-0.19} \times 10^{-5}$ & $1.26^{+0.15}_{-0.15}
\times 10^{-5}$ \\
& $F_{\rm narrow}\,(\ergpcmsqps)$ & $1.00^{+0.16}_{-0.16} \times 10^{-13}$ & $9.54^{+1.54}_{-1.54}
\times 10^{-14}$ & $1.23^{+0.18}_{-0.18} \times 10^{-13}$ \\
& $EW_{\rm narrow}\,(\eV)$ & $111^{+18}_{-18}$ & $111^{+18}_{-18}$ & $121^{+16}_{-16}$ \\
\hline
{\tt laor}, {\tt kdblur} & $\alpha_{\rm emis}$ & $2.40^{+2.14}_{-0.82}$ & $2.40^{+2.17}_{-0.65}$ & $0.95^{+9.05}_{-0.95}$ \\
& $r_{\rm in}\,(r_{\rm g})$ & $19.93^{+24.52}_{-10.04}$ & $19.76^{+24.68}_{-19.76}$ & $20.15^{+25.36}_{-20.15}$ \\
& $i(\degmark)$ & $72^{+0}_{-15}$ & $72^{+0}_{-15}$ & $72^{+0}_{-15}$ \\
& $K_{\rm broad}\,(\phpcmsqps)$ & $1.69^{+1.03}_{-0.69} \times 10^{-5}$
  & $1.71^{+0.97}_{-0.68} \times 10^{-5}$ & --- \\
& $F_{\rm broad}\,(\ergpcmsqps)$ & $1.44^{+0.50}_{-0.59} \times 10^{-13}$ & $1.38^{+0.78}_{-0.55} \times
10^{-13}$ & --- \\
& $EW_{\rm broad}\,(\eV)$ & $185^{+112}_{-75}$ & $187^{+106}_{-74}$ & --- \\
\hline
{\tt pexrav}, {\tt reflion} & ${\rm Fe/solar}$ & --- & $1.0$ & $1.0$ \\
& $\xi\,(\ergpcmps)$ & --- & $0.0$ & $109^{+2}_{-109}$ \\
& $R_{\rm refl}$ & --- & $0.01^{+0}_{-0.01}$ & $5.75^{+0}_{-5.75} \times 10^{-3}$ \\
& $K_{\rm refl}\,(\phpcmsqps)$ & --- & $7.21^{+0}_{-7.21} \times
10^{-8}$ & $1.30^{+1.71}_{-1.22} \times 10^{-7}$ \\
& $F_{\rm refl}\,(\ergpcmsqps)$ & --- & $4.04^{+0}_{-4.04} \times 10^{-18}$ & $4.24
\times 10^{-13}$ \\
\hline
& $\chi^2/\nu$ & $660/685\,(0.96)$ & $660/683\,(0.97)$ & $660/684\,(0.96)$ \\
\hline \hline
\end{tabular}
\end{center}
\caption{\small Comparison of our three best-fit models for the $0.5-35 \keV$
  spectrum of NGC~1052.  
All components of Models~1-3 are modified by Galactic hydrogen
  absorption with $N_{\rm H}=2.83 \times 10^{20} \pcmsq$.  
``K'' denotes the normalization value of a
  given component.  Flux values indicate absorbed flux from
  $0.5-35 \keV$.  Redshifts for the model components are held
  constant at the cosmological value for NGC~1052: $z=0.0049$.
  Abundances not listed are held constant at their solar values.  The
  inclination angle of the accretion disk used in the {\tt laor} and
  {\tt kdblur} models
  was constrained to fall within the radio observation uncertainties
  of $i=57-72\degmark$.  All
  errors listed are at $90\%$ confidence for one interesting parameter.}
\label{tab:tab1}
\end{table}

\begin{sidewaystable}
\begin{center}
\begin{tabular}{|l|l|l|l|}
\hline \hline
{\bf Waveband (keV)} & {\bf Model~1 Flux ($\ergpcmsqps$)} & {\bf Model~2
  Flux ($\ergpcmsqps$) } & {\bf Model~3 Flux ($\ergpcmsqps$)} \\
\hline
$0.5-2.0$ & $3.42 \times 10^{-13}$ & $3.42 \times 10^{-13}$ & $3.68
\times 10^{-13}$ \\
 & $3.42 \times 10^{-12}$ & $3.42 \times 10^{-12}$ & $5.31 \times 10^{-12}$ \\ 
\hline
$2.0-10.0$ & $5.53 \times 10^{-12}$ & $5.25 \times 10^{-12}$ & $5.33
\times 10^{-12}$ \\
 & $8.96 \times 10^{-12}$ & $8.96 \times 10^{-12}$ & $9.20 \times 10^{-12}$ \\
\hline
$10.0-60.0$ & $3.64 \times 10^{-11}$ & $3.64 \times 10^{-11}$ & $3.79
\times 10^{-11}$ \\
 & $3.68 \times 10^{-11}$ & $3.68 \times 10^{-11}$ & $3.82 \times 10^{-11}$ \\
\hline \hline
\end{tabular}
\end{center}
\caption{\small Extrapolated model fluxes for the 2007 {\it Suzaku} observation of NGC~1052 in
  three wavebands.  Absorbed fluxes are on the top line, unabsorbed
  fluxes are on the line below.  For the softest energies the flux from XIS~1 was
  used, while the XIS~0+3 data was used for the $2-10 \keV$ band.  PIN
  data were used for the flux above $10 \keV$.}
\label{tab:tab2}
\end{sidewaystable}

\section{Comparison with Previous Results}
\label{sec:epoch}

Here we consider our best 2007 {\it Suzaku} results (Model~1) against those obtained in the
1996 {\it ASCA} observation by W99 and the 2000 {\it BeppoSAX}
observation by G00.  We do not elect to discuss individually either the 1999 {\it
  ROSAT} results from \citet{Guainazzi1999} nor the 2000 {\it
  Chandra} results from \citet{Kadler2004b} due to the inferior
signal-to-noise of these observations. 
 
Neither W99 nor G00 presented a light curve for their data, but both
authors failed to detect any significant variability over the course
of their observations.  W99 fit a time-averaged
$2-10 \keV$ flux of $F_{2-10} \sim 8 \times 10^{-12} \ergpcmsqps$,
corrected for a dual absorber in the best fit.  
This differs only slightly from the unabsorbed flux of $F_{2-10} \sim
9 \times 10^{-12} \ergpcmsqps$ we have obtained from our data in the
present epoch.  G00 use only one cold intrinsic absorber in their best fit,
which yields an unabsorbed flux of $F_{2-10} \sim 9 \times 10^{-12}
\ergpcmsqps$, also consistent with our 2007 observation. 

The neutral hydrogen absorbing column in our best fit is less dense
than that of W99 or G00.
We find evidence for one intrinsic neutral absorber of $N_{\rm H} \sim 1.08^{+0.04}_{-0.08}
\times 10^{23} \pcmsq$ (with a partial-covering fraction of $\sim
83^{+3}_{-4}\%$).  In their best-fitting model, W99
employ a two-zone intrinsic absorption model with $N_{\rm H}(1)=3.00^{+1.68}_{-1.16} \times 10^{23}
\pcmsq$ and $N_{\rm H}(2)=4.9^{+2.0}_{-1.4} \times 10^{22} \pcmsq$.  The G00 best-fit
model uses one intrinsic absorber of $N_{\rm H}=2^{+0.6}_{-0.5} \times 10^{23}
\pcmsq$.   Our model
contains a component of partly-ionized absorption that W99 and G00 do
not, however: the {\tt absori} model representing this absorbing
material has a column density of $N_{\rm H}=1.37^{+0.54}_{-0.78}
\times 10^{22} \pcmsq$
and an ionization parameter of $\xi=68^{+100}_{-31} \ergpcmps$.  Even with this
additional layer the W99 and G00 models contain higher absorbing columns than
our 2007 model, however.  

The heavy absorption in W99, in particular, resulted in a power-law photon index
split between intrinsic and scattered components, where a power-law
photon index of $\Gamma=1.7$ was assumed and
$11\%$ was scattered, while $70\%$ of the intrinsic flux was absorbed by
the larger column and $30\%$ by the smaller column.  We attempted
such a neutral dual-absorber fit (see \S\ref{sec:xis}-\ref{sec:fe}) and 
found that the parameters of the second absorber could not be
constrained, and that a single intrinsic neutral absorber and an
ionized absorber along with Galactic
photoabsorption modifying a simple, albeit fairly hard power-law ($\Gamma \sim 1.5$)
provided an excellent fit to the continuum.  

W99 and G00 also found evidence for a thermal {\tt mekal} component
to explain the soft excess noted below $2 \keV$, which was best
modeled with $kT=0.53^{+0.34}_{-0.26} \keV$ and a $0.1-2.0 \keV$ flux of $F_{\rm
  mekal}=5.8^{+2.5}_{-2.3} \times 10^{-14} \ergpcmsqps$ in W99.  
G00 modeled this component with a similar model of much higher
temperature of $kT \geq 5 \keV$, though these authors do not report its flux.
This thermal
component was also detected by \citet{Kadler2004b} with {\it Chandra}
at $kT=0.41^{+0.09}_{-0.07}$, though over a flux range of $0.2-8.0
\keV$ and with limited signal-to-noise.
We found a similar thermal component in the {\it Suzaku} data.  Though the
XIS detectors lose effective area rapidly below $0.3 \keV$,
extrapolating Model~1 down to $0.1 \keV$ yielded
$kT=0.64 \pm 0.04 \keV$ and $F_{\rm mekal}=9.38 ^{+2.27}_{-2.31} \times 10^{-14} \ergpcmsqps$
over $0.1-2.0 \keV$.  Our results showed a slight increase in
temperature and flux for the {\tt mekal} component but were consistent
with the 1996 results within error bars.

In the favored dual-absorber {\it ASCA} model of W99,
a broad Fe K$\alpha$ line component was not required in the fit and
hence these authors found
no definitive evidence for reflection from an optically-thick
accretion disk in their data.  They did require a narrow Fe K$\alpha$
line, however: the centroid energy of this component was held constant at $E=6.37
\keV$ and had an equivalent width of $EW=270 \pm 120 \eV$.  G00 did
detect an Fe K line with some apparent broadening, though this could
be accounted for in the {\it BeppoSAX} data with a mildly ionized Fe
K$\alpha$ line with centroid energy $E=6.48^{+0.16}_{-0.20} \keV$ and
an equivalent width of $EW=230 \pm 170 \eV$.
In our Model~1 for the {\it Suzaku} data, we found that both broad and
narrow components were necessary in order to eliminate the residuals
between $4-7 \keV$.  Our narrow line, with a centroid energy 
of $6.4 \pm 0.01 \keV$ had an
equivalent width of $111 \pm 18 \eV$.  The presence of a
broad component to the Fe K$\alpha$ line likely rendered this narrow feature
significantly weaker than its 1996 counterpart.  

Our broad Fe K$\alpha$ line, held constant at a centroid energy of $E=6.4 \keV$,
was best fit with a {\tt laor} emission feature originating
from the accretion disk around a rapidly rotating black hole, though
the inner radius of emission within the disk was only mildly
constrained to $r_{\rm in}<45\,r_{\rm g}$, as
discussed in \S\ref{sec:fe}.  This feature was stronger than the
narrow Fe K$\alpha$ line: its equivalent
width was $185^{+112}_{-75} \eV$.  Replacing the {\tt laor} line with
a broad Gaussian, we found that $\sigma=1.04 \pm 0.26 \keV$, which translates to
a lower-limit FWHM of $v \sim 0.37 c$ and implies an origin for this feature very
close to the black hole. 
Though the fit of the {\tt laor} line itself is suggestive of a broad, fluorescent
Fe K$\alpha$ origin in the inner accretion disk, the lack of reflected
continuum flux above $10 \keV$ argues against this interpretation, as
we discuss below and in the following Section.

W99 considered a reflection-based {\tt hrefl} model to
fit the {\it ASCA} spectrum, but were unable to obtain meaningful
constraints on the reflection parameters without making several
assumptions.  These included fixing the value of the power-law photon
index of the central X-ray source to be $\Gamma=1.7$ and also fixing
the inclination angle of the disk to $i=60\degmark$.  The latter was
consistent with our own angle constraints based on the VLBI observations
of NGC~1052, but our {\it Suzaku} observation indicated a
somewhat harder photon index (see Table~\ref{tab:tab1}).
For an assumed reflection fraction of $R_{\rm refl}=1$ and the
observed {\it ASCA} spectrum, W99 found that only a small
fraction of the direct emission from the X-ray source was visible,
leading these authors to infer that a thick absorbing column in the nucleus
($N_{\rm H} > 3 \times 10^{23} \pcmsq$) blocked the X-ray source from
view, such that reflected emission dominated the spectrum.  However,
the equivalent width of the Fe K$\alpha$ line, $EW \sim 40 \eV$, was
over an order of magnitude smaller than the same line seen in other
heavily obscured AGN.  To account for this line with a reflection
model would require an implausibly small iron abundance of $Z_{\rm
  Fe}/{\rm solar} \leq 0.05$.  This reflection model was therefore
argued against by W99.  Similarly, G00 also ruled out evidence for
Compton reflection in the NGC~1052 {\it BeppoSAX} spectrum due to a
statistically unacceptable fit for the {\tt pexriv} model.  These
authors were only able to mildly constrain the reflection fraction to
a hard upper limit of $R_{\rm refl}<0.6$.

The 2007 {\it Suzaku} data concur with this
assessment of a lack of reflection: though Model~2 produced a comparable global
goodness-of-fit to Model~1, the reflection fraction was found to be
quite small: $R_{\rm refl}<0.01$.  Attempting to model any reflection present
with Model~3, we were unable
to constrain the emissivity index of the accretion
disk, though we did achieve similar constraints for the equivalent
width of the narrow iron line and the photon index of the power-law component.
Based on this spectral fitting, we conclude that in 2007, 
NGC~1052 does show evidence for a broadened iron line; this line is not particularly broad or
strong, however, and is not accompanied by the expected Compton
hump above $10 \keV$.  This may indicate that the line arises outside of the
innermost portions of the disk and/or that reflection simply does not
play a large role in creating the overall X-ray spectrum. 

\begin{figure}
\begin{center}
\includegraphics[width=0.6\textwidth,angle=270]{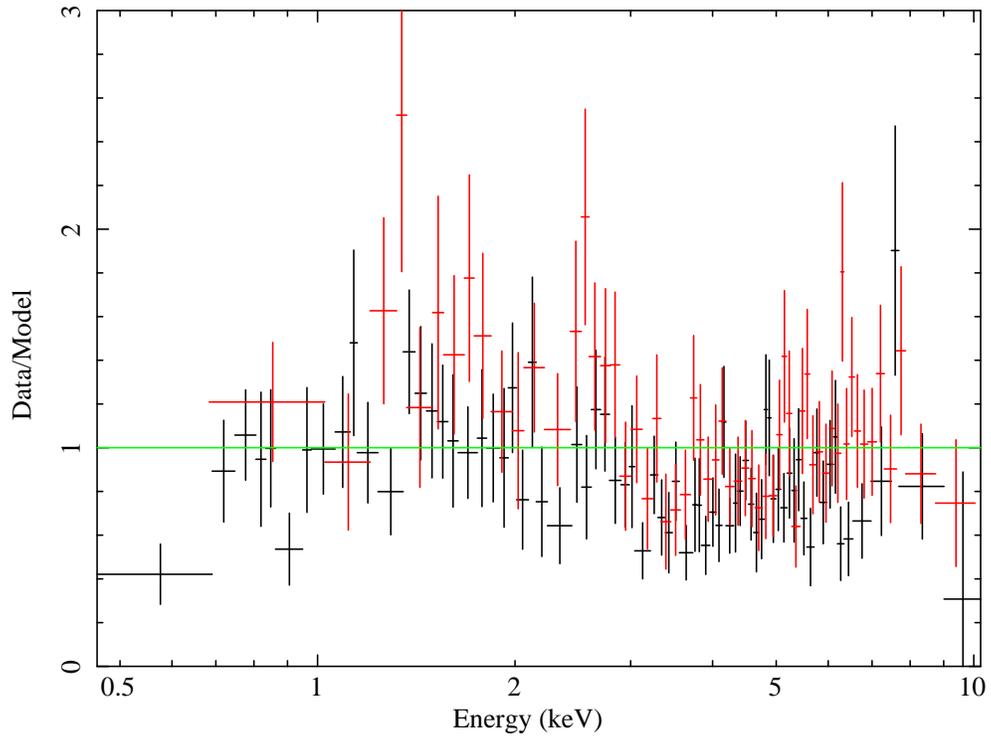}
\end{center}
\caption{\small Ratio plot of the 1996 {\it ASCA} spectrum of NGC~1052 to the
2007 {\it Suzaku} best-fit Model~1.  SIS~0 and GIS~2 data are shown with
black and red crosses, respectively.  The green line denotes a
theoretical perfect fit.  Note the curvature of the residuals below
$\sim 3 \keV$.  These residuals indicate that changes have
occurred in the absorption
structure surrounding the AGN between epochs.
$\chi^2/\nu=255/141\,(1.81)$.  No refit has taken place.}
\label{fig:asca_suzmo}
\end{figure}

In order to make a more detailed comparison between the 1996 {\it
ASCA} data and the 2007 {\it Suzaku} data for NGC~1052, we have
examined the $0.5-10 \keV$ W99 spectrum in combination with our
Model~1.  Data from the SIS~0 and GIS~2 were used, as these were
the instruments with the highest count rates and the closest on-axis
telescope pointing during the {\it ASCA} observation.  We show a plot
of the ratio of the {\it ASCA} data to our {\it Suzaku} best-fit
Model~1 in Fig.~\ref{fig:asca_suzmo}.  

Changes between the two epochs are evident
at energies below $\sim 3 \keV$ and above $\sim 7 \keV$, likely
indicating that the continuum and absorbing structures have  undergone physical
changes during the intervening years.  Refitting the model, we
found that $\chi^2/\nu=120/141\,(0.85)$ as opposed to
$\chi^2/\nu=255/141\,(1.81)$ before refitting.  Reduced chi-squared
values less than unity often indicate that the model has more
parameters than are necessary to adequately fit the data, and given
the lower spectral resolution and throughput of the {\it ASCA} data,
this is likely the case here.  Nonetheless, we can use this
refit to verify which model components have undergone significant
changes between 1996 and 2007.  

The thermal
component had a temperature of $kT=0.50 \pm 0.14 \keV$, consistent
with the results of W99 and consistent in its upper limit with
Model~1.  The normalizations of the thermal
component were also consistent within error bars.  The power-law
photon index from Model~1 was $\Gamma=1.50^{+0.07}_{-0.08}$,
vs. $\Gamma=1.28 \pm 0.91$ from our refit of the {\it ASCA} data, also
consistent within errors.  W99
had maintained this component at a fixed value of $\Gamma=1.7$ for
their best-fitting models, but our refitted value was consistent with
the W99 fits for their models~2-4.  Though W99 employed a power-law component
split between intrinsic and scattered emission while we determined
that only intrinsic emission was necessary for the {\it Suzaku} data,
the total normalization of all the W99 power-law components was
consistent with our intrinsic power-law normalization within errors.
Our refitted {\it ASCA} value (assuming intrinsic emission only) was
also consistent with the W99 intrinsic normalization within errors.
Refitting the absorption parameters yielded an ionized absorbing column
of $N_{\rm H} \leq 5.64 \times 10^{21}\pcmsq$ and an ionization
parameter of $\xi \leq 97 \ergcmps$.  While these values were not
well-constrained, they were at least marginally consistent with those
of Model~1.  However, while the partial-covering absorber column was
constrained to $N_{\rm H} \leq 15.19 \times 10^{22} \pcmsq$,
consistent with Model~1, the covering fraction could not be
constrained.  These changes suggest that the absorbing structure
differs physically in 2007 as compared with 1996. 

The iron line complex, by contrast, showed little variation between the two epochs.  Our refit
to the {\it ASCA} data yielded a narrow Fe K$\alpha$ component of normalization
$K=1.19^{+0.83}_{-0.78} \times 10^{-5} \phpcmsqps$, consistent with
our {\it Suzaku} Model~1 value and the W99 fit, and strongly
indicating that this component, if variable, is only variable on long
timescales and must therefore be emitted from material far from the black hole.  W99
reported that no broad Fe K$\alpha$ line component was robustly present in 1996,
contrary to our findings in 2007.  Our refit, including a {\tt laor}
component, does not adequately constrain the parameters of the broad
line in 1996, confirming the null result of W99.  This could mean 
either that the broad line component was truly absent in 1996, or perhaps
that it was present at a lower normalization and effectively drowned
out by the continuum emission.  Given the rapid decrease in effective
area above $\sim 5 \keV$ and the spectral resolution of {\it ASCA},
however, it is very likely that the broad iron line component simply could
not have been detected by this observatory, thus rendering moot any
comparison between a broad Fe K$\alpha$ line in 1996 and 2007.

\section{Discussion}
\label{sec:discussion}

We have performed detailed spectral fits to the XIS and HXD/PIN data
from the $101 \ks$, 2007 {\it Suzaku} observation of the LINER galaxy
NGC~1052.  Three different spectral models were employed: (1) an
absorbed power-law continuum with contributions from a soft thermal
component along with narrow and broad Fe K$\alpha$ lines;
(2) the same model, but with an added component of reflected continuum
emission from a neutral accretion disk ({\tt pexrav}); (3) the same
base model, again, but with a more advanced, self-consistent
reflection model which includes fluorescent emission lines from the
disk and takes ionization into account ({\tt reflion}; see
  \citet{Ross2005} for details).  In Model~3 we applied relativistic smearing
using the {\tt kdblur} model.  Because fluorescent lines are
included in {\tt reflion}, there is no need for an additional broad
iron line component as in Model~1.  We summarize our findings below:

\begin{itemize}
\item{The three models are quite similar in their statistical
  goodness-of-fit, though some of the disk reflection parameters in
  Models~2-3 prove
  especially difficult to constrain.}
\item{The continuum is well described by a component with a power-law
  spectrum of photon index $\Gamma \sim
  1.5$, thermal emission with $kT \sim 0.64 \keV$, a neutral intrinsic
  absorbing column of $N_{\rm H} \sim 10^{23} \pcmsq$ and a
  covering fraction of $\sim 84\%$, and an ionized intrinsic absorber
  with $N_{\rm H} \sim 1.37 \times 10^{22} \pcmsq$ and $\xi \sim 68 \ergpcmps$.
  These findings are relatively
  consistent with those of the 1996 {\it ASCA} observation of W99,
  though our photon index is harder and our absorbing column
  lower in density in 2007.}
\item{As in the {\it ASCA} data, a narrow Fe K$\alpha$ line is
  present: $EW \sim 111 \eV$.}
\item{Strong evidence
  exists for a broadened Fe K$\alpha$ emission component ($EW \sim 185
  \eV$).  Though the best statistical fit is achieved with a {\tt
  laor} line profile, however,
  the rather broad constraint on the inner radius of $r_{\rm
  in}<45\,r_{\rm g}$ renders it impossible to distinguish between a
  spinning and non-spinning black hole.}
\item{The PIN instrument did detect hard X-ray emission above $10
  \keV$ from NGC~1052.  This emission is well fit by the power-law
  component of the $2-10 \keV$ continuum and does not show any evidence for
  reflected emission via the 
  Compton hump that is commonly associated with other AGN harboring
  broad iron lines, and expected based on the presence of the broad
  iron line if the two features arise from reflection onto a standard
  accretion disk.  Neither of our reflection models found any
  statistical evidence for a strong reflection component.}
\item{Though VLBI observations of the jet in NGC~1052 have constrained
the orientation angle of the disk to be between $i=57-72\degmark$, we
find that removing this constraint in our fitting yields a best-fit
inclination angle of $i=45 \pm 5 \degmark$, though this does not
significantly improve the global goodness-of-fit.  Nonetheless, this
result does mark a $\sim 2\sigma$ deviation from the radio results.}
\end{itemize}

These findings motivate three main questions: (1) what is the physical
origin of the broad iron line in our 2007 observation; (2) how is it
that we detect a broad iron line yet no evidence for disk reflection
above $10 \keV$, and (3) why is the {\tt laor} line inclination,
when kept as a free parameter in the fit, not more consistent with the disk
inclination constraints established in radio observations of the inner
jet in NGC~1052? 

As it turns out, the answers to these questions may be closely related.
Relativistically broad iron lines are expected to be associated with pronounced Compton
humps.  Comptonized reflection originates
via irradiation of the optically thick inner accretion disk by the hard X-ray source,
perhaps in some sort of lamp-post geometry, e.g., the base of a jet
\citep{Miniutti2007}.
This same process will also produce fluorescent line emission from the
disk, most notably the Fe K$\alpha$ line, which will experience
significant broadening and skewness by virtue of its origin from the
spacetime proximal 
to the black hole \citep{R+N2003}.  The absence of a Compton hump,
however, casts doubt on the presence/contribution of
reflection from the disk to the overall X-ray spectrum.  From our {\tt
  pexrav} model fit (Model~3), we can estimate the equivalent width of the iron
line we should expect if both the Compton hump and the broad line
originate from disk reflection.  Using the ratio of the normalization
of the power-law to the upper limit of the reflected component in our fit, we infer that the
iron line equivalent width should be $EW \sim 80 \eV$ if the two
features are indeed produced by the same physical process.  This low value
is highly inconsistent with the observed strength of the broad iron
feature, however.

Relativistic disk smearing
is not the only process by which this spectral line can be broadened.
In the following paragraphs we consider a number of other physical scenarios, including: 
transition to an advection-dominated accretion flow (ADAF) in the
inner disk, resonant scattering from the putative torus region, origin
of the line in an outflow located in the broad line region (BLR), and
Fe K$\alpha$ produced by Cerenkov line-like radiation rather than
fluorescence.   

One explanation for the lack of an observed reflected continuum could
be that the accretion disk transitions to a radiatively-inefficient
state within some radius, e.g., \citet{Narayan1994}.  In this ADAF
picture, which has often been invoked as a potential explanation for
low-luminosity sources (especially those with observed jet activity),
the disk traps radiation rather than releasing
it, puffing up to become optically thin and geometrically thick as its
temperature and ionization state rises.  Because of these properties,
the gas in this region would not be expected to produce much, if any, reflection,
either in the form of a Compton hump or discrete emission lines.  If a
contribution to the
broad iron line did somehow arise from this region, we would expect it to be
highly ionized, yet this is not seen in the data: our broad and narrow
iron line components both have energies robustly less than $6.41
\keV$, suggesting that they arise from species at or below Fe\,{\sc
  xviii}.  We therefore conclude that an ADAF, if present, is unlikely
to be a significant contributor to the broadening of the Fe K$\alpha$ feature.

The lack of a Compton hump at the expected strength and the mismatch between the
unconstrained {\tt laor} disk
inclination and the VLBI value causes us to rule out a relativistic
inner disk origin for the broad iron line.  We also eliminate a possible
origin in an ADAF due to ionization constraints and optical depth
issues.  Investigating other possible sources of iron emission in the
central AGN leads us to consider the role of resonant scattering in
the torus region, which is thought to contribute heavily to the flux
of the narrow Fe K$\alpha$ line.  However,
both the large equivalent width and the large FWHM velocity of the
broad line component we observe (both substantially greater than the
$v \sim 2500 \kmps$ and $EW \lesssim 100 \eV$ typically associated
with the narrow core, e.g., \citet{Yaqoob2004}) render this
explanation implausible.  

An AGN outflow originating in the BLR is explored as a possible
explanation for the similar spectrum of NGC~7213 \citep{Bianchi2008}.
Here a broad iron line with no corresponding Compton hump is also
observed, and these authors find that the equivalent width of the
Fe K$\alpha$ feature is consistent with predictions made by
\citet{Yaqoob2001} for an origin in the BLR or perhaps a Compton-thin
torus.  Further, measurements of the velocity broadening of the
optical H$\alpha$ line concur with those measured from the broad
Fe K$\alpha$ line ($v=2500 \kmps$), suggesting a similar origin for
the emission.  Using the same diagnostic, we estimate that we should expect
a broad line $EW \sim 75-100 \eV$ if it originates in an outflow from
this region; this underestimates our observed equivalent width by
nearly a factor of ten using a simple Gaussian line to represent the
broad feature, and by roughly a factor of two if the better-fitting
(though, in this case, unphysical) {\tt laor} line is used.  Also, our
estimated lower-limit FWHM velocity of the
line ($v \sim 0.37c$) is more than an order of magnitude greater than
the polarized optical H$\alpha$ FWHM measurement of 
\citet{Barth1999} ($v \sim 5000 \kmps$). This is a strong indication
that the 
broad Fe K$\alpha$ line we observe originates closer to the
black hole than the BLR.  Moreover, variability of the broad iron line on
years-long timescales has been noted \citep{Kadler2004c,Ros2007}; this type of
variation would not be expected from the typical symmetric BLR, as
changes in flux due to passing clouds in this region should
theoretically average out over time.

Finally, we consider a more exotic possibility: perhaps the broad iron
line is produced by Cerenkov line-like radiation in the centralmost
regions of the AGN.  \citet{You2003} put forward the intriguing notion
that in dense regions where the refractive index of the material is
large, relativistic electrons impinging upon this material can produce
Cerenkov radiation in a narrow wavelength range very close to the
intrinsic atomic wavelength of the material.  The combination of
absorption and emission causes the final emission feature to be
redshifted and skewed, making it appear very like a relativistic 
line produced by fluorescence in the inner accretion disk.
Because electrons are producing the emission rather than
photons, no reflected continuum is expected.  The cool, dense, iron-rich gas
in the disk of NGC~1052 could provide the medium for this process, while 
relativistic electrons in the corona or near the base of the jet would make for
an ideal bombardment population.  Furthermore, this physical picture makes
for a much more consistent match with our estimate of the distance of
the broad line region of origin from the hard X-ray source: our
lower-limit FWHM velocity for the broad Gaussian fit to the line
yields $v \sim 0.37c$, which corresponds to an upper-limit distance of $d \sim
8\,r_{\rm g}$.  This is well inside the BLR.

Given our constraint on the distance of the broad iron line emitter from the hard
X-ray source, iron fluorescence from the base of the jet itself is also a
potential candidate for the emission mechanism.  However, considering
the orientation of the jets observed in NGC~1052 to our line of sight,
it is unclear whether any fluorescent emission from the jet base
could escape the system without being absorbed.  Correlated radio flux
measurements from the base of the jet and
X-ray spectra of this source must be examined over several
epochs, simultaneously whenever possible, in order to assess the
likelihood of this
scenario as well as that of the Cerenkov line-like radiation proposed
by \citet{You2003}.  Only through a coordinated study of the jet and the
accretion flow in NGC~1052 can we hope to understand the connection
between these two vital physical processes in AGN.

\section{Conclusions}
\label{sec:concl}

Our 2007 {\it Suzaku} spectrum of the LINER galaxy NGC~1052 is
consistent with X-ray spectra of this source from previous epochs,
with a fairly flat power-law continuum that is heavily absorbed.
Intrinsic neutral and ionized columns are detected, along with
evidence of Galactic photoabsorption.  A thermal component
is also detected, likely due to the interaction of the jets with the
surrounding ISM.  Both narrow and broad iron lines are observed,
though interestingly there is little to no reflection seen in the
spectrum above $10 \keV$: two different models used to characterize
this feature both require $R_{\rm refl}<0.01$.

We are thus faced with a complex scenario for NGC~1052 in which the
broad iron line is not consistent with the lack of observed continuum reflection
from the disk.  While it is possible that the reflection has been
drowned out in some way by the power-law continuum, we must also
consider the physical implications if reflection is indeed absent
above $10 \keV$.  It could be that we are witnessing the transition of
the accretion flow to a radiatively-inefficient state at some critical
radius in the disk, which could largely eliminate reflection features from the
spectrum.  A broad iron line is still produced, however, and the
velocity width of the line is consistent with an origin close to the
black hole.  If the inner disk is an ADAF, it is possible that the
broad Fe K$\alpha$ line is emitted from optically-thick material near the base of the jet,
or perhaps from a yet more exotic mechanism such as the Cerenkov-like
line radiation postulated by \citet{You2003}.  Coordinated
observations of both the inner jet(s) (radio) and the inner accretion
flow (X-ray) are needed in order to solve this puzzle.

\section*{Acknowledgments}

LB and MK thank
the NASA Postdoctoral Program, administered by ORAU, for their 
support at NASA's GSFC.  KW gratefully acknowledges support from NASA
grant NNX08AC226, MA from NASA grant NNX08AC22G.
YYK is a Research Fellow of the Alexander von Humboldt Foundation.
LB and KW also appreciate excellent advice
from Tahir Yaqoob, Chris Reynolds, Julian Krolik, Demos Kazanas and Tim Kallman.
  KW would
like to dedicate this work to Professor Andrew Wilson, who prompted her
interest in LINERs and NGC~1052 in the 1990s.  We gratefully
acknowledge the helpful comments received from our anonymous referee,
which have improved this work.

\bibliographystyle{apj}
\bibliography{adsrefs}

\end{document}